\documentclass[twocolumn]{aastex62}
\usepackage{color,soul}
\usepackage{amsmath}

\defcitealias{Kennicutt98}{K98}
\defcitealias{RC3}{RC3}
\defcitealias{PT05}{PT05}

\newcommand{\galex}{\textit{GALEX}}
\newcommand{\xco}{$X(\textrm{CO})$}
\newcommand{\molgas}{$\textrm{H}_{2}$}

\newcommand{\mips}{\textit{MIPS}}
\newcommand{\iras}{\textit{IRAS}}
\newcommand{\wise}{\textit{WISE}}

\graphicspath{{./}{figures/}}

\accepted{December 20, 2018}
\submitjournal{ApJ}

\shorttitle{Paper I: The Integrated Star Formation Law}
\shortauthors{de los Reyes et al.}

\begin{document}

\title{Revisiting the Integrated Star Formation Law. Paper I: Non-Starbursting Galaxies}

\correspondingauthor{Mithi A. C. de los Reyes}
\email{mdelosre@caltech.edu}

\author[0000-0002-4739-046X]{Mithi A. C. de los Reyes}
\affiliation{Department of Astronomy, California Institute of Technology \\
1200 E. California Blvd., MC 249-17 \\
Pasadena, CA 91125, USA }
\affiliation{Institute of Astronomy, University of Cambridge \\
Madingley Road, Cambridge CB3 0HA, UK}

\author{Robert C. Kennicutt Jr.}
\affiliation{Department of Astronomy and Steward Observatory, University of Arizona \\
Tucson, AZ 85721, USA}
\affiliation{Department of Physics and Astronomy, Texas A\&M University \\
College Station, TX 77843, USA}
\affiliation{Institute of Astronomy, University of Cambridge \\
Madingley Road, Cambridge CB3 0HA, UK}

\begin{abstract}

We use new and updated gas and dust-corrected SFR surface densities to revisit the integrated star formation law for local ``quiescent'' spiral, dwarf, and low-surface-brightness galaxies. 
Using UV-based SFRs with individual IR-based dust corrections, we find that ``normal'' spiral galaxies alone define a tight $\Sigma_{(\textrm{HI}+\textrm{H}_{2})}$-$\Sigma_{\textrm{SFR}}$ relation described by a $n=1.41^{+0.07}_{-0.07}$ power law with a dispersion of $0.28^{+0.02}_{-0.02}$ (errors reflect fitting and statistical uncertainties).
The SFR surface densities are only weakly correlated with HI surface densities alone, but exhibit a stronger and roughly linear correlation with H$_{2}$ surface densities, similar to what is seen in spatially-resolved measurements of disks.
However, many dwarf galaxies lie below the star formation law defined by spirals, suggesting a low-density threshold in the integrated star formation law.
We consider alternative scaling laws that better describe both spirals and dwarfs.
Our improved measurement precision also allows us to determine that much of the scatter in the star formation law is intrinsic, and we search for correlations between this intrinsic scatter and secondary physical parameters.
We find that dwarf galaxies exhibit second-order correlations with total gas fraction, stellar mass surface density, and dynamical time that may explain much of the scatter in the star formation law.
Finally, we discuss various systematic uncertainties that should be kept in mind when interpreting any study of the star formation law, particularly the $X(\textrm{CO})$ conversion factor and the diameter chosen to define the star-forming disk in a galaxy.

\end{abstract}


\section{Introduction} 
\label{sec:intro}

The formation of stars from the interstellar medium (ISM) is one of the driving processes in galaxy evolution.
Although the global star formation rate (SFR) of a galaxy is likely set by a balance of complex physical processes, the nature of the overall relationship between SFR and the ISM can be parameterized by simple empirical scaling laws.
One such law is the relationship between gas density and star formation rate density, known as the star formation law or Schmidt law \citep{Schmidt59}.
The Schmidt law was originally posed as a power law relationship between volume densities, but \citet{Schmidt63} recast it as a relation between surface densities $\Sigma$:
\begin{equation}
\Sigma_{\textrm{SFR}} = A(\Sigma_{\textrm{gas}})^{n},
\label{eq:SchmidtLaw}
\end{equation} 
where both quantities are integrated measurements of global galaxy properties.

\citet{Kennicutt98} (hereafter \citetalias{Kennicutt98}) found that quiescent spiral galaxies, infrared-luminous starbursts, and circumnuclear starbursts obeyed a tight relationship defined by Equation~\ref{eq:SchmidtLaw} with power-law index $n=1.4\pm 0.15$. 
(Here, a ``starburst'' is defined as a system with an SFR much higher than the long-term average SFR of the system; in comparison, a ``quiescent'' galaxy has an SFR roughly consistent with an equilibrium system.)
This Schmidt law may imply either that the local density of gas drives star formation efficiency $\epsilon\equiv\Sigma_{\textrm{SFR}}/\Sigma_{\textrm{gas}}$ in a ``bottom-up'' formulation, or that star formation is driven by ``top-down'' dynamical processes \citep{Kennicutt12rev}.
Regardless of its physical interpretation, this law has been widely applied as a recipe for star formation in cosmological simulations, many of which lack the spatial resolution needed to model complex sub-grid physics.
\citetalias{Kennicutt98} also found that quiescent spirals and starbursts obeyed an alternative version of the Schmidt law, posed by \citet{Silk97} and \citet{Elmegreen97}:
\begin{equation}
\Sigma_{\textrm{SFR}} = A\frac{\Sigma_{\textrm{gas}}}{\tau_{\textrm{dyn}}},
\label{eq:SELaw}
\end{equation} 
where $\tau_{\textrm{dyn}}$ represents the dynamical (orbital) timescale \citepalias{Kennicutt98}.
This version may imply that global dynamical features like spiral arms or bars might convert a constant fraction of gas into stars. 
This concept was subsequently generalized by \citet{Krumholz2012} into a universal correlation between SFR surface density and gas surface density per free fall time.

Since 1998, observational efforts have shifted to studying the star formation law on the scale of star-forming regions \textit{within} individual galaxies.
These spatially-resolved studies have presented a range of results. 
On global scales, \citetalias{Kennicutt98} found that SFR correlates strongly with total gas surface density, moderately with atomic gas surface density ($\Sigma_{\textrm{HI}}$), and only weakly with molecular gas surface density ($\Sigma_{\textrm{H}_{2}}$).
Most of the spatially-resolved studies, on the other hand, find that SFR correlates most strongly with $\Sigma_{\textrm{H}_{2}}$ and almost not at all with $\Sigma_{\textrm{HI}}$ \citep[e.g.,][]{Kennicutt07,Bigiel08,101,Bigiel14}.

Furthermore, many spatially-resolved studies report a shallower star formation law slope $n$ than that found by \citetalias{Kennicutt98}.
These works also suggest that the molecular gas depletion time ($\tau_{\textrm{depl}}(\textrm{H}_{2})\equiv\Sigma_{\textrm{H}_{2}}/\Sigma_{\textrm{SFR}}$) is constant \citep[e.g.,][]{Bigiel08, 225, Leroy13}.
However, studies of larger samples show that a larger range of $\tau_{\textrm{depl}}(\textrm{H}_{2})$ may instead depend systematically on specific SFR \citep[SSFR$\equiv$SFR$/M_{*}$;][]{Saintonge11b}.
Other studies also find a steeper $\Sigma_{\textrm{SFR}}$-$\Sigma_{\textrm{H}_{2}}$ law \citep{Kennicutt07}, which may be a result of removing the ``diffuse'' local infrared and ultraviolet background \citep{Liu11,Momose13,MorokumaMatsui17}.

While the spatially-resolved star formation law may shed more light on the local physical processes driving star formation, the global star formation law still plays a vital role. 
In many scenarios (e.g., high-redshift galaxies) only global measurements are available.
Furthermore, spatially-resolved SFRs are subject to more physical uncertainty than globally averaged SFRs.
This is due to the stochastic nature of star formation on local scales.

For example, SFR tracers are generally sensitive to the high-mass end of the stellar initial mass function (IMF), but the IMF is often poorly sampled within small regions \citep[i.e., on spatial scales of $\sim 0.1-1$~kpc in typical star-forming galaxies;][]{Kennicutt12rev}, leading to large variations in tracer luminosities for a given SFR.
Indirect SFR tracers, such as H$\alpha$ and infrared luminosities, can also be biased by emission from diffuse gas and dust, which may be located far from actual regions of star formation; these can affect measurements on the scale of hundreds of parsecs \citep{Kennicutt12rev}.
Finally, stellar ages can fluctuate dramatically on small spatial scales (i.e., spatial scales small enough to be dominated by very young stellar clusters). This produces uncertainty since SFR is computed as the mass of recently-formed stars divided by the time over which they were formed.
In part because of these complicated systematic uncertainties, it is important to understand the scale dependences (if any) in the star formation law.

Indeed, investigations of the global law since 1998 have raised additional questions about the nature of star formation.
For instance, low surface brightness galaxies show evidence for a turnover in the star formation law at low gas surface densities \citep{Wyder09}. 
At the high-density end, infrared-luminous starburst galaxies may define a star formation law that bifurcates from that for normal disk galaxies \citep{Genzel10,Daddi10}, though this may depend on the treatment of the CO-to-\molgas{} conversion factor \xco{} \citep{Narayanan12}.
Other scaling laws have also been proposed, including relationships between SFR and \textit{dense} gas mass \citep{Gao04} or between SFR and a combination of gas and stellar surface densities \citep[e.g.,][]{Shi11,Kim15}.

These developments motivate a fresh investigation of the global star formation law. 
\citet{Liu15} recently re-analyzed the global Schmidt law using 1.4~GHz radio continuum sizes and SFR measurements.
This analysis confirmed basic results from \citetalias{Kennicutt98} but found a significantly shallower power-law slope $n$.
The cause of this discrepancy is not immediately clear, and understanding it is another motivation for this study.

To investigate the issues raised above and test the conclusions from \citetalias{Kennicutt98}, we aim to update the global star formation law with a larger sample and more accurate data.
Improvements in available multi-wavelength data now make it possible to carry out a more comprehensive analysis.
In particular, the increased availability of spatially-resolved HI maps and CO data provides gas surface density measurements for more galaxies.
Similarly, the Galaxy Evolution Explorer (\galex{}) provides ultraviolet (UV) fluxes that can be used to compute SFRs for much larger samples of galaxies.
The UV measurements also allow us to extend reliable SFR measurements to dwarf galaxies and other systems with low SFRs, where H$\alpha$ rates are subject to large uncertainties introduced by stochasticity in the instantaneous SFR \citep[e.g.,][]{Lee09}.
With these spatially-resolved maps, we are able to define physically-based radii for averaging surface densities, rather than arbitrarily using optical or radio continuum isophotes.
These larger samples also enable us to significantly extend the range of galaxy types and surface densities probed by the star formation law.
Finally, infrared (IR) measurements and new prescriptions for dust attenuation corrections make it possible to improve the precision of SFR measurements, allowing us to search for secondary physical parameters driving the dispersion in the relation.

We present the results in two papers.
In this paper, we revisit the integrated star formation law for non-starbursting spiral, dwarf, and low surface brightness galaxies. 
We aim to determine if spiral galaxies alone can define a tight correlation between gas and SFR surface densities. 
By extending the surface density range probed to over three orders of magnitude, we also address other questions about the low-density regime of global star formation in galaxies.
Paper II (Kennicutt \& de los Reyes, in prep.) considers starbursts and high surface density systems, as well as the combined relation over all densities.

This paper is organized as follows.
In Section~\ref{sec:data}, we present the dataset used in this sample.
Our main results are outlined in Section~\ref{sec:results}, and we consider the possibility of second-order correlations in the star formation law in Section~\ref{sec:secondorder}. 
We discuss the limitations of our dataset in Section~\ref{sec:uncertainties} before considering literature comparisons (Section~\ref{sec:litcompare}) and the physical implications of our results (Section~\ref{sec:interpretations}). Finally, we summarize our findings in Section~\ref{sec:summary}.

\section{Data} 
\label{sec:data}
In this section, we present the multi-wavelength data used to measure star formation rate densities\footnote{Unless otherwise noted, we henceforth use ``densities'' to refer to ``surface densities.''}, gas densities, and other galaxy properties.

\subsection{Sample selection}
\label{sec:sample}
Our base sample is composed of $N=307$ nearby galaxies with good coverage in UV, mid-IR, and radio wavelengths.
In particular, these galaxies were selected based on the availability of CO maps and spatially-resolved HI maps.
Known luminous active galactic nuclei (AGN) were removed from the sample to prevent AGN radiation from being misidentified as radiation from star formation.

\begin{figure}
	\epsscale{1.2}
	\plotone{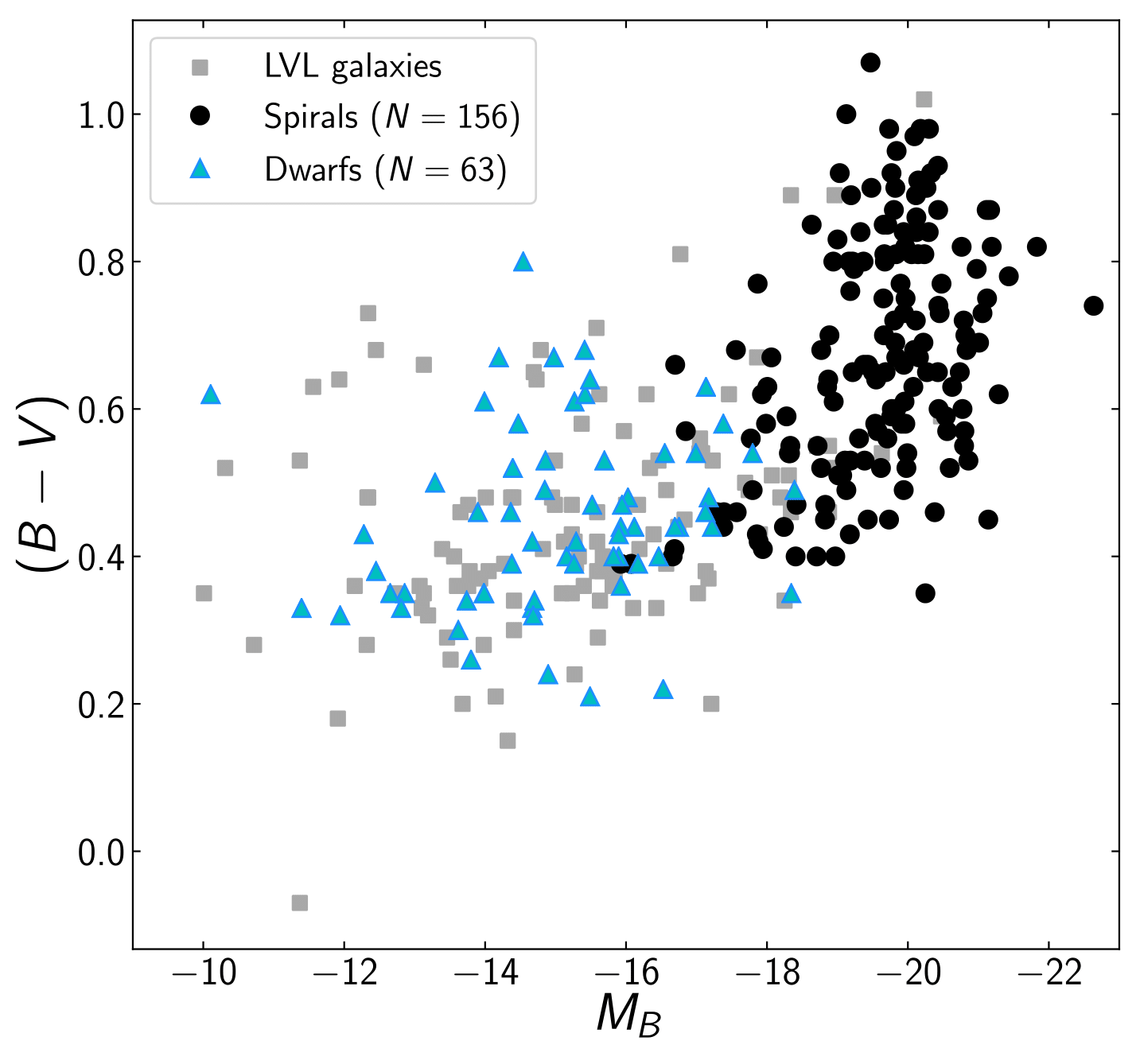}
	\caption{Color-magnitude diagram of the spiral (black solid points) and dwarf (cyan triangles) galaxies in our sample. Local Volume Legacy (LVL) galaxies not included in our sample are plotted as small gray points for comparison.\label{fig:dwarfspiralclasses}}
\end{figure}

To characterize our sample, we plot a color-magnitude diagram of our sample in Figure~\ref{fig:dwarfspiralclasses}.
We also plot the Local Volume Legacy \citep[LVL;][]{dale09} galaxies, a volume-limited sample of galaxies within 11~Mpc of the Milky Way, on Figure~\ref{fig:dwarfspiralclasses} (small gray points) for comparison.
The overlap between our sample and the LVL sample suggests that our galaxies form a generally representative sample of local galaxies, suitable for studying the general star formation law.

Of the $N=307$ galaxies in the sample, $169$ are typical star-forming disk or ``spiral'' galaxies, while the remaining $138$ are ``dwarf'' galaxies. 
We initially defined dwarf galaxies as galaxies with low stellar masses ($M_{*}\leq 10^{9}$~$M_{\odot}$) or low luminosities ($M> -17$~mag).
However, the exact definition of dwarf galaxies is somewhat ambiguous, and we manually reclassified several dwarf galaxies---particularly low-luminosity disks---based on clear morphological distinctions.
We also removed known blue compact dwarf (BCD) galaxies from our sample, and defer discussion of these highly-starbursting systems to Paper II.
This reclassification does not significantly affect the results of any part of our analysis; as shown in Figure~\ref{fig:dwarfspiralclasses}, the final ``spiral'' and ``dwarf'' populations are still largely distinct on a color-magnitude diagram.
We note that the spirals have a relatively narrow range in physical properties, as is typical of the blue sequence.

\begin{deluxetable*}{lllllllllllll}
\tablecolumns{13} 
\tablecaption{ General properties of sample. \label{tab:gendata}} 
\tablehead{ 
\colhead{N} & \colhead{NED ID} & \colhead{R.A.} & \colhead{Dec} & \colhead{Dist} & \colhead{$m_{B}$} & \colhead{B-V} & \colhead{Type} & \colhead{$\theta$} & \colhead{E(B-V)} & \colhead{$D_{25}$} & \colhead{$D_{\textrm{H}\alpha}$} & \colhead{$b/a$} \\
\colhead{} & \colhead{} & \colhead{(J2000)} & \colhead{(J2000)} & \colhead{(Mpc)} & \colhead{(mag)} & \colhead{(mag)} & \colhead{} & \colhead{(deg)} & \colhead{} & \colhead{(\arcsec)} &  \colhead{(\arcsec)} & \colhead{}
}
\startdata
1 & WLM & 00h01m58.16s & -15d27m39.3s & 0.92 & \ldots & $0.44 \pm 0.04$ & d & 4 & $0.04$ & 689 & 154 & 0.35 \\
2 & NGC 7817 & 00h03m58.91s & +20d45m08.4s & 26.13 & $12.56 \pm 0.18$ & \ldots & s & 45 & $0.06$ & 213 & 105 & 0.26 \\
3 & NGC 0023 & 00h09m53.41s & +25d55m25.6s & 56.22 & $12.56 \pm 0.14$ & $0.82 \pm 0.05$ & s & 8 & $0.04$ & 125 & 55 & 0.65 \\
4 & UGC 00191 & 00h20m05.20s & +10d52m48.0s & 15.9 & $14.89 \pm 0.26$ & $0.44 \pm 0.04$ & d & 150 & $0.11$ & 97 & 70 & 0.74 \\
5 & M 031 & 00h42m44.35s & +41d16m08.6s & 0.79 & $4.16 \pm 0.19$ & $0.92 \pm 0.02$ & s & 35 & $0.58$ & 11433 & 5940 & 0.32 \\
6 & IC 1574 & 00h43m03.82s & -22d14m48.8s & 4.92 & $14.47 \pm 0.21$ & $0.61 \pm 0.04$ & d & 0 & $0.02$ & 128 & 60 & 0.36 \\
7 & NGC 0253 & 00h47m33.12s & -25d17m17.6s & 3.94 & $8.27 \pm 0.21$ & $0.85 \pm 0.05$ & s & 52 & $0.02$ & 1653 & 1200 & 0.25 \\
8 & UGCA 015 & 00h49m49.20s & -21d00m54.0s & 3.34 & $15.34 \pm 0.21$ & $0.43 \pm 0.05$ & d & 42 & $0.02$ & 102 & 100 & 0.43 \\
9 & NGC 0278 & 00h52m04.31s & +47d33m01.8s & 11.8 & $11.49 \pm 0.22$ & $0.64 \pm 0.01$ & s & 30 & $0.14$ & 125 & 45 & 0.95 \\
10 & UGC 00634 & 01h01m25.10s & +07d37m35.0s & 27.2 & $15 \pm 0.3$ & $0.48 \pm 0.06$ & d & 35 & $0.05$ & 100 & 88 & 0.65
\enddata
\tablecomments{Only a portion of Table~\ref{tab:gendata} is shown here; it is published in its entirety in the machine-readable format online.}
\end{deluxetable*}

Table~\ref{tab:gendata} presents the basic properties of this sample. Columns 1-8 list general properties of the sample, while columns 9-12 include data used to perform photometry (Sections~\ref{sec:SFRdata} and \ref{sec:gasdata}).

Column 1: running index number.

Column 2: galaxy name preferred by the NASA/IPAC Extragalactic Database (NED)\footnote{The NASA/IPAC Extragalactic Database (NED) is operated by the Jet Propulsion Laboratory, California Institute of Technology, under contract with the National Aeronautics and Space Administration.}.

Columns 3 and 4: J2000 right ascension and declination as reported in NED.

Column 5: distance, adopted from the 11~Mpc H$\alpha$ UV Galaxy Survey \citep[11HUGS;][]{Kennicutt08} catalog when possible or from NED otherwise.

Column 6: apparent B-band AB magnitude, from \citet{RC3} (hereafter \citetalias{RC3})

Column 7: B-V color, from \citetalias{RC3}.

Column 8: galaxy classification as a ``dwarf'' (d) or ``spiral'' (s).

Column 9: position angle. 

Column 10: reddening E(B-V), adopted from \citet{schlegel98} unless otherwise noted.

Column 11: $D_{25}$ diameter, defined as the major axis of the \citetalias{RC3} $B$-band 25~mag~arcsec$^{-2}$ isophote.

Column 12: $D_{\textrm{H}\alpha}$ diameter, defined as the semimajor axis of the region containing ${\sim}95\%$ of the H$\alpha$ flux (Section~\ref{sec:diams}).

Column 13: ratio of semi-minor to semi-major axis, measured from B-band isophotes by \citetalias{RC3}.

\subsection{Diameters}
\label{sec:diams}
Since we aim to obtain SFR and gas surface densities rather than total SFRs and gas masses for the star formation law, we must define a diameter
\footnote{Note that the choice of a single star-forming diameter is not necessarily the most physically correct choice. For a variety of reasons (e.g., extended gas distributions affecting star formation on more compact scales), it is certainly possible to define different gas and SFR diameters, or even to plot total gas masses and SFRs (i.e., total efficiencies). However, a full discussion of diameter choice is beyond the scope of this work. Here, we choose instead to consider an ``idealized'' case of a single star-forming region to measure both gas and SFR densities, and we defer deeper discussion to Paper II.} 
by which to normalize our measured quantities.
Rather than using an optically-defined size such as the $D_{25}$ isophotal diameter, we choose to define a \textit{star-forming region} as the region containing ${\sim}95\%$ of the H$\alpha$ flux (Table~\ref{tab:gendata}).
This is largely motivated by an attempt to remain self-consistent with our work with starburst galaxies (Paper II).
In these galaxies, nearly all of the star formation is confined to circumnuclear molecular disks on the scale of $1$~kpc, an order of magnitude smaller than the scale of the optical disk; the $D_{25}$ diameter is therefore not a relevant scale for studying global star formation in these galaxies.

The H$\alpha$ flux, on the other hand, is a more direct tracer of star formation than the near-IR. 
It is also less susceptible to dust attenuation than other star formation tracers, such as the UV continuum (see next section).
Finally, H$\alpha$ maps of local galaxies are readily available from the literature, including the 11HUGS survey \citep{Kennicutt08}. 
However, we note that the use of H$\alpha$-based diameters may produce uncertainties for some galaxies; for example, in some cases extended UV emission can be found without obvious H$\alpha$ counterparts \citep[e.g.,][]{Thilker07,Goddard10}.
We note that the H$\alpha$-defined diameter is generally more compact than both the $D_{25}$ diameter and a UV-defined diameter, with $D_{25} = 1.83 D_{\textrm{H}\alpha}$ and $D_{\textrm{UV}} = 2.24D_{\textrm{H}\alpha}$ on average.
In Section~\ref{sec:uncertainties}, we consider potential systematic effects on our analysis that may arise from defining the star-forming region by H$\alpha$ flux.

\subsection{SFR surface densities}
\label{sec:SFRdata}
The most physically direct measure of SFR is the ultraviolet continuum produced by young stars, which traces SFR within the past ${\sim}10-200$~Myr.
The UV continuum is less sensitive to fluctuations in the high-mass end of the stellar initial mass function (IMF) than other SFR tracers, particularly emission-line tracers such as H$\alpha$, which has been shown to be unreliable in low-SFR regions \citep{Lee09}.
We therefore use UV-based SFRs throughout our analysis.

The main disadvantage of UV light as a SFR tracer is its sensitivity to dust extinction and reddening.
Since dust absorbs starlight at all wavelengths and re-emits in the infrared, IR luminosities can be used to correct for this attenuation \citep[e.g.,][]{Hao11}.

In this section, we describe the UV and IR photometry obtained for our sample, as well as the SFR and SFR density calculations.
The final SFR and SFR densities are listed in Table~\ref{tab:photometry} at the end of the section.

\subsubsection{UV photometry}
UV data were obtained by the Galaxy Evolution Explorer (\galex{}), a $50$ cm aperture space telescope launched in 2003 by NASA. 
\galex{} consisted of a 50~cm aperture space telescope that took simultaneous FUV ($\lambda_{\textrm{eff}}=1539$\AA) and NUV ($\lambda_{\textrm{eff}}=2316$\AA) observations. 
For this work, FUV is preferentially used; older stars contribute to the near-UV flux, so NUV-based SFR calibrations are more sensitive to the recent star formation history and the assumed stellar IMF.

When possible, we use \galex{} FUV fluxes from various literature catalogs.
In order of priority, we compiled FUV aperture fluxes from: the \citet{GildePaz2007} Atlas of Nearby Galaxies; the Local Volume Legacy (LVL) survey \citep{lee11} \citep[using apertures matched to the IR apertures of][]{dale09}; the Spitzer Infrared Nearby Galaxies Survey \citep[SINGS;][]{dale07}; the LVL survey \citep{lee11} (using apertures defined by the ``outermost elliptical annulus where both FUV and NUV photometry can be performed''); \citet{Bai15}, the Virgo Cluster Survey \citep{Voyer14}; and the Herschel Reference Survey \citep{Cortese12}.

Eighty galaxies in the base $N=307$ sample were not included in the above catalogs.
Of these $80$ galaxies, $68$ had available \galex{} imaging, and we performed aperture photometry for most of these using the deepest available \galex{} FUV images. 
For $12$ galaxies, FUV observations were not available, and we used the deepest available NUV images instead.
The photometric procedure is described in detail in Appendix~\ref{appendix:photometry}.
As described in Appendix~\ref{appendix:photcorrections}, we also performed FUV photometry for an additional $59$ galaxies to check for consistency between our measurements and the catalogs. 
We find that most of the discrepancies between our measured fluxes and the catalog fluxes arise from differences in aperture size, since the literature catalogs report either asymptotic fluxes or aperture fluxes with different aperture sizes.
Statistical aperture correction factors, tabulated in Table~\ref{tab:apcorr}, were therefore applied to the catalog fluxes to correct for this aperture effect.

The final UV fluxes---either aperture-corrected catalog fluxes, or our measured photometric fluxes---are listed in Table~\ref{tab:UVdata}.
Table~\ref{tab:UVdata} also summarizes the observations of the $139$ galaxies for which we measured photometric fluxes.

Column 1: running index number.

Column 2: galaxy name preferred by NED.

Column 3: UV apparent magnitude.

Column 4: reference for UV flux.

Column 5: exposure time.

Column 6: \galex{} tile number.

Column 7: NUV flag (0 if FUV flux was available, 1 if FUV unavailable and NUV used instead).

All images used for photometric measurements were preprocessed using the latest available \galex{} pipeline \citep{martin05,morrissey05,morrissey07}.

\begin{deluxetable*}{cllcllc}
\tablecolumns{7} 
\tablecaption{ UV photometric data. \label{tab:UVdata}} 
\tablehead{ 
\colhead{N} & \colhead{NED ID} &  \colhead{$m_{\textrm{UV}}$} & \colhead{UV reference} & \colhead{$t_{\textrm{exp}}$} & \colhead{Tile} & \colhead{NUV flag} \\
\colhead{} & \colhead{} & \colhead{(mag)} & \colhead{} & \colhead{(s)} & \colhead{} & \colhead{}
}
\startdata
1 & WLM & 12.8 $\pm$ 0.1 & 2 & \ldots & \ldots & \ldots \\
2 & NGC 7817 & 15.0 $\pm$ 0.2 & 1 & 90. & AIS\_144 & 1 \\
3 & NGC 0023 & 16.4 $\pm$ 0.4 & 1 & 3398.05 & GI1\_013001\_NGC0023 & 0 \\
4 & UGC 00191 & \ldots & \ldots & \ldots & \ldots & \ldots \\
5 & M 031 & 8.4 $\pm$ 0.1 & 6 & \ldots & \ldots & \ldots \\
6 & IC 1574 & 16.9 $\pm$ 0.1 & 3 & \ldots & \ldots & \ldots \\
7 & NGC 0253 & 11.6 $\pm$ 0.1 & 2 & \ldots & \ldots & \ldots \\
8 & UGCA 015 & 17.1 $\pm$ 0.1 & 3 & \ldots & \ldots & \ldots \\
9 & NGC 0278 & 11.8 $\pm$ 0.1 & 1 & 106. & AIS\_43 & 1 \\
10 & UGC 00634 & 16.6 $\pm$ 0.5 & 1 & 197.05 & AIS\_264 & 0
\enddata
\tablecomments{Only a portion of Table~\ref{tab:UVdata} is shown here; it is published in its entirety in the machine-readable format online.}
\tablereferences{(1) This paper; (2) \citet{GildePaz2007}; (3) LVL, using IR-matched apertures \citep{lee11}; (4) SINGS \citep{dale07}; (5) LVL, using ``outermost elliptical aperture'' \citep{lee11}; (6) \citet{Bai15}; (7) Virgo Cluster Survey \citep{Voyer14}; (8) Herschel Reference Survey \citep{Cortese12}.}
\end{deluxetable*}

\subsubsection{IR photometry}
\label{sec:IRphotometry}
The IR data in this work come from three instruments: the Spitzer Space Telescope's Multiband Imaging Photometer (\mips{}), the Infrared Astronomical Satellite (\iras{}), and the Wide-field Infrared Survey Explorer (\wise{}).
Spitzer \mips{}, launched in 2003 by NASA, contains separate detector arrays that perform imaging and spectroscopy at 24, 70, and 160~$\mu$m bands; the 24~$\mu$m images, used in this work, were taken by a camera with a 5' square field of view \citep{Rieke04}.
\iras{}, launched in 1983 as a joint project between the US, the UK, and the Netherlands, performed an all-sky survey at 12, 25, 60, and 100~$\mu$m \citep{Neugebauer84}.
Finally, \wise{} was a NASA space telescope launched in 2009; the data in this paper were obtained during the original \emph{4-Band (or Full) Cryogenic survey}, which observed the entire sky in 3.4, 4.6, 12 and 22~$\mu$m bandpasses \citep{Wright10}.

Although dust re-radiates attenuated UV light at all wavelengths in the IR \citep{Draine03}, complete wavelength coverage is available for few galaxies.
We therefore use data from the $\approx$24~$\mu$m band, since Spitzer \mips{}, \iras{}, and \wise{} all have comparable bandpasses, and monochromatic dust corrections at this wavelength are only slightly less accurate than corrections using total IR \citep{Hao11}.
In Section~\ref{sec:uncertainties}, we discuss potential uncertainties arising from our use of a single infrared band rather than total IR.

When possible we use mid-IR fluxes from existing catalogs.
In order of priority, we compiled either Spitzer \mips{} 24\micron{} or \iras{} 25\micron{} fluxes from: the LVL survey \citep{dale09}; SINGS \citep{dale07}; the \citet{GildePaz2007} Atlas of Nearby Galaxies; the \mips{} Local Galaxy Survey \citep[\mips{} LG;][]{Bendo12}; and the \iras{} Revised Bright Galaxies Sample \citep[\iras{} BGS;][]{Sanders03}.
As before, we also performed photometry for $46$ galaxies using Spitzer \mips{} images to check for consistency between our measurements and the catalog fluxes.
These images were pre-processed by the LVL, SINGS, and \mips{} LG surveys \citep{dale09, dale07, Bendo12}; the processing steps (including instrumental correction, calibration, and background subtraction) are described in detail in the above references.
We find that the catalog fluxes must be corrected to account for not only differences in aperture size, but also differences in the bandpass wavelength (24\micron{} for Spitzer \mips{} or 25\micron{} for \iras{}).
Both the photometry procedure and the correction factors are described in Appendix~\ref{appendix:photometry}.

The above catalogs do not contain $104$ of the galaxies in our sample.
We performed photometry for these galaxies using 22\micron{} data from \wise{}'s AllWISE Image Atlas\footnote{The AllWISE explanatory supplement can be found at \url{http://wise2.ipac.caltech.edu/docs/release/allwise/expsup/}}.
The AllWISE images used for photometry come from the \wise{} Image Atlas, which are co-adds of corrected \wise{} frames. 
As described in Appendix~\ref{appendix:photometry}, we also performed additional photometry using AllWISE data to compare with the fluxes we measured from Spitzer \mips{} images.
A statistical correction factor is applied to the AllWISE fluxes to correct for the difference in AllWISE 22\micron{} and Spitzer 24\micron{} bandpasses, as well as for variations in background subtraction methods.

The final IR fluxes---either aperture-corrected catalog fluxes, or our measured AllWISE fluxes---for our sample are listed in Table~\ref{tab:IRdata}.
Table~\ref{tab:IRdata} also summarizes the IR observations of the $150$ galaxies for which we measured photometric fluxes.

Column 1: running index number.

Column 2: galaxy name preferred by NED.

Column 3: 24\micron{} infrared flux.

Column 4: reference for IR flux.

Column 5: exposure time.

Column 6: for Spitzer \mips{} images, the name of the survey that pre-processed the image.

Column 7: for AllWISE images, the number of frames (single-band images).

Column 8: for AllWISE images, the co-add (image produced by combining single-exposure frames) ID.

\begin{deluxetable*}{cllcllll}
\tablecolumns{8} 
\tablecaption{ IR photometric data. \label{tab:IRdata}} 
\tablehead{ 
\colhead{N} & \colhead{NED ID} &  \colhead{$f_{\textrm{IR}}$} & \colhead{IR reference} & \colhead{$t_{\textrm{exp}}$$^{\textrm{a}}$} & \colhead{Project} & \colhead{$n_{\textrm{frames}}$} & \colhead{Co-add ID} \\
\colhead{} & \colhead{} & \colhead{(Jy)} & \colhead{} & \colhead{(s)} & \colhead{} & \colhead{} & \colhead{}
}
\startdata
1 & WLM & (70 $\pm$ 9)$\times 10^{-3}$ & 3 & \ldots & \ldots & \ldots & \ldots \\
2 & NGC 7817 & (29 $\pm$ 2)$\times 10^{-2}$ & 2 & \ldots & \ldots & 99 & 0016p212\_ac51 \\
3 & NGC 0023 & (11 $\pm$ 2)$\times 10^{-1}$ & 7 & \ldots & \ldots & \ldots & \ldots \\
4 & UGC 00191 & (4 $\pm$ 2)$\times 10^{-3}$ & 2 & \ldots & \ldots & 124 & 0046p106\_ac51 \\
5 & M 031 & 98 $\pm$ 23 & 5 & \ldots & \ldots & \ldots & \ldots \\
6 & IC 1574 & $<6.87\times 10^{-4}$ & 2 & \ldots & \ldots & 150 & 0114m228\_ac51 \\
7 & NGC 0253 & 139 $\pm$ 17 & 3 & \ldots & \ldots & \ldots & \ldots \\
8 & UGCA 015 & $<1\times 10^{-3}$ & 2 & \ldots & \ldots & 141 & 0129m213\_ac51 \\
9 & NGC 0278 & (22 $\pm$ 5)$\times 10^{-1}$ & 7 & \ldots & \ldots & \ldots & \ldots \\
10 & UGC 00634 & (3 $\pm$ 2)$\times 10^{-3}$ & 2 & \ldots & \ldots & 127 & 0152p075\_ac51
\enddata
\tablenotetext{a}{\mips{} LG files do not list exposure times. See \citet{Bendo12} for more details.}
\tablecomments{Only a portion of Table~\ref{tab:IRdata} is shown here; it is published in its entirety in the machine-readable format online. }
\tablereferences{(1) This paper (using Spitzer MIPS images); (2) This paper (using AllWISE images); (3) LVL \citep{dale09}; (4) SINGS \citep{dale07}; (5) \citet{GildePaz2007}; (6) \mips{} LG \citep{Bendo12}; (7) \iras{} BGS \citep{Sanders03}. }
\end{deluxetable*}

\subsubsection{SFRs and SFR surface densities}
\label{sec:SFRcalibs}
After converting UV and IR fluxes to luminosities using the distances reported in Table~\ref{tab:gendata}, the UV luminosity can be corrected for dust correction by using an energy balance argument \citep{Hao11}:
\begin{align}
L(\textrm{FUV})_{\textrm{corr}} & = L(\textrm{FUV})_{\textrm{obs}} + (3.89\pm 0.20)L(24\mu\textrm{m})_{\textrm{obs}}\\
L(\textrm{NUV})_{\textrm{corr}} & = L(\textrm{NUV})_{\textrm{obs}} + (2.26\pm 0.16)L(24\mu\textrm{m})_{\textrm{obs}}
\label{eq:dust}
\end{align}

The SFR is then calculated using the calibrations of \citet{Murphy11}, assuming a Kroupa IMF \citep{Kroupa01}.
These calibrations are also tabulated in \citet{Kennicutt12rev}:
\begin{align}
\log[\textrm{SFR}~(\textrm{M}_{\odot}~\textrm{yr}^{-1})] & = \log[L(\textrm{FUV})_{\textrm{corr}}~(\textrm{erg}~\textrm{s}^{-1})] - 43.35\\
\log[\textrm{SFR}~(\textrm{M}_{\odot}~\textrm{yr}^{-1})] & = \log[L(\textrm{NUV})_{\textrm{corr}}~(\textrm{erg}~\textrm{s}^{-1})] - 43.17
\label{eq:SFR}
\end{align}
These calibrations are appropriate for estimating SFRs integrated over entire galaxies; other works have considered SFR estimation on smaller spatial scales \citep[e.g.,][]{Leroy12}.

Finally, the total SFR is converted to SFR surface density $\Sigma_{\textrm{SFR}}$ by normalizing by the de-projected area of the star-forming region $\pi R^{2}$.
As noted in Section~\ref{sec:diams}, this star-forming region is defined as the region containing ${\sim}95\%$ of the H$\alpha$ flux, so that the physical radius $R$ is computed from the distance to the galaxy and the semi-major axis $a$ of the star-forming region.
Table~\ref{tab:gendata} lists these diameters, and Table~\ref{tab:photometry} lists the SFR surface densities.

We defer a discussion of potential systematic uncertainties arising from these SFR calculations, including the choice of diameter and the recipe for dust correction, to Section~\ref{sec:uncertainties}.

\subsection{Gas surface densities}
\label{sec:gasdata}
Disk-averaged surface densities of HI and H$_{2}$ were compiled from published 21\,cm and CO measurements in the literature, and it is the availability of these data which primarily determines the selection of galaxies for this study.
Unless otherwise stated, the surface densities quoted are for hydrogen alone; densities including helium can be derived by multiplying by a factor of $1.36$.

It is well known that the HI disks of galaxies extend in most cases well beyond the main star-forming disks, and the total HI masses and surface densities averaged over the entire HI disk often deviate significantly from the mean densities in the star-forming regions.  
Consequently, following \citetalias{Kennicutt98} we restricted our interest to galaxies with well-resolved HI maps, usually measured from aperture synthesis arrays, in order to measure the mean HI surface density over the same physical region as for the molecular gas and the SFR.  
Data were compiled from 114 papers as listed in Table~\ref{tab:photometry}, though a majority of the data come from a handful of large surveys made with the Westerbork Synthesis Radio Telescope (WSRT), the Very Large Array (VLA), or the Giant Millimetre Radio Telescope (GMRT).
In most cases the primary papers present azimuthally-averaged HI surface density profiles, which were used to determine the average surface density $\Sigma_{\textrm{HI}}$ within the radius of the star-forming region, listed in Table~\ref{tab:gendata}; otherwise, these were derived from the published contour maps.
Care was taken to exclude galaxies where the beam size was too large to determine an accurate average column density.  
In a handful of instances (mostly dwarf galaxies), we used a single-dish flux and star-forming radius to estimate the mean HI surface density, but only in cases where resolved maps were unavailable and the relevant beam size was comparable to the diameter of the star-forming disk.  
All surface densities were deprojected to face-on orientation.

Mean molecular hydrogen surface densities $\Sigma_{\textrm{H}_{2}}$ were calculated from published CO(1-0) and/or CO(2-1) measurements. 
For the galaxies in this paper, most of these data were obtained with single-dish millimeter telescopes using single-beam (often with multiple pointings across the disks) or multi-beam arrays.  
In nearly all cases, the CO emission is restricted to a region of size comparable to or smaller than the star-forming disks, so in practice we adopted the published fluxes and molecular gas masses and divided the latter by the de-projected area of the star-forming region to determine the
mean surface density.  
Care was taken to correct these values to a common CO to H$_2$ conversion factor (see below), assumed $\textrm{CO}(2-1)/\textrm{CO}(1-0)$ ratio, and a molecular hydrogen (only) mass.  
When data from multiple sources were available, they were averaged.  
Many of these measurements consisted of a series of pointings (ususally along the major axis), and the integrated fluxes were computed from the resulting fitted radial profile of CO emission. 
We adopted these published values, but excluded galaxies with insufficient radial coverage for a reliable estimate of the total flux.  
This interpolation is often the dominant source of uncertainty in the fluxes ($\pm30\%$).

Estimates of molecular hydrogen masses derived from CO rotational line measurements are notorious for their dependence on a variable CO/H$_2$ conversion factor \citep[$X(\textrm{CO})$ or $\alpha$(CO); see][]{Bolatto13}.
Both the average value adopted and the prescriptions for parametrizing systematic variability in $X(\textrm{CO})$ have evolved considerably since \citetalias{Kennicutt98}.
For most of the subsequent analysis we chose to adopt a constant value:
\begin{equation}
X(\textrm{CO}) =  2.0 \times 10^{20}~\textrm{cm}^{-2}~(\textrm{K~km~s}^{-1})^{-1},
\label{eq:XCO_MW}
\end{equation}
which appears to apply to the molecular disks of most quiescent spirals \citep{Bolatto13}.  
However, this value almost certainly does not apply in low-mass dwarf galaxies, and when we discuss that subsample in Section~\ref{sec:results}, we will consider alternative formulations for
$X(\textrm{CO})$.

The uncertainties in the mean surface densities for HI and H$_2$ listed in Table~\ref{tab:photometry} are dominated by the signal/noise of the maps (HI and CO), and corrections for spatial undersampling of the disks (CO).  
Since the surface density measurements come from a variety of sources, we assume a conservative estimate of measurement uncertainty: 0.1 dex $(\sim26\%)$ each for $\log \Sigma_{\textrm{HI}}$ and $\log \Sigma_{\textrm{H}_{2}}$.
Uncertainties in the adopted radii also propagate into the SFRs, but these are relatively small for HI (which tend to have flatter radial profiles), and for H$_2$ any deviation will be identical to that in the mean SFR surface density.
Other systematic uncertainties in the surface densities not included in Table~\ref{tab:photometry} will be discussed in Section~\ref{sec:uncertainties}.

\begin{deluxetable*}{llrrccccc}
\tablecolumns{11} 
\tablecaption{SFRs, gas masses, and SFR and gas surface densities. \label{tab:photometry}} 
\tablehead{ 
\colhead{N} & \colhead{NED ID} & \colhead{$\log \textrm{SFR}_{\textrm{uncorr}}$} & \colhead{$\log \textrm{SFR}_{\textrm{corr}}$} & \colhead{$\log \Sigma_{\textrm{SFR}}$} & \colhead{$\log \Sigma_{\textrm{HI}}$$^{\textrm{a}}$} & \colhead{HI ref.}  & \colhead{$\log \Sigma_{\textrm{H}_{2}}$$^{\textrm{a}}$} & \colhead{H$_{2}$ ref.}\\
\colhead{} & \colhead{} & \colhead{($[M_{\odot} \textrm{yr}^{-1}]$)} & \colhead{($[M_{\odot} \textrm{yr}^{-1}]$)} & \colhead{($[M_{\odot} \textrm{yr}^{-1} \textrm{kpc}^{-2}]$)} & \colhead{([$M_{\odot} \textrm{pc}^{-2}$])} & \colhead{} & \colhead{([$M_{\odot} \textrm{pc}^{-2}$])} & \colhead{}
}
\startdata
1 & WLM & $-2.59\pm0.04$ & $-3.74\pm0.06$ & -3.31 & 0.81 & 127 & \ldots & \ldots \\
2 & NGC 7817 & $-0.14\pm0.04$ & $-0.21\pm0.03$ & -2.35 & 0.94 & 8 & 1.06 & 2 \\
3 & NGC 0023 & $0.96\pm0.09$ & $1.02\pm0.09$ & -1.23 & 0.66 & 106 & 1.70 & 2 \\
4 & UGC 00191 & \ldots & $-2.49\pm0.21$ & -3.85 & 0.93 & 146 & \ldots & \ldots \\
5 & MESSIER 031 & $-0.59\pm0.07$ & $-0.73\pm0.10$ & -3.34 & 0.31 & 102 & -0.38 & 3,201 \\
6 & IC 1574 & $<$-2.79 & $<$-4.29 & -4.50 & 0.49 & 149 & \ldots & \ldots \\
7 & NGC 0253 & $0.76\pm0.06$ & $0.82\pm0.05$ & -1.80 & 0.59 & 104,105 & 0.89 & 1,202,228 \\
8 & UGCA 015 & $<$-3.20 & $<$-4.40 & -4.71 & 0.41 & 149,154 & \ldots & \ldots \\
9 & NGC 0278 & $0.24\pm0.04$ & $-0.02\pm0.09$ & -0.74 & 1.00 & 106 & 2.10 & 1 \\
10 & UGC 00634 & $-1.17\pm0.18$ & $-2.24\pm0.43$ & -4.27 & 0.94 & 146 & \ldots & \ldots
\enddata
\tablenotetext{a}{As noted in the text, we assume conservative measurement uncertainties of $\pm 0.1$~dex for $\log \Sigma_{\textrm{HI}}$ and $\log \Sigma_{\textrm{H}_{2}}$.}
\tablecomments{Only a portion of Table~\ref{tab:photometry} is shown here; it is published in its entirety in the machine-readable format online.}
\tablereferences{See Appendix~\ref{appendix:gasrefs}}
\end{deluxetable*}

Table~\ref{tab:photometry} lists the computed SFRs, gas masses, and SFR and gas surface densities, as follows.

Column 1: running index number.

Column 2: galaxy name preferred by NED.

Column 3: total (UV-based) SFR, uncorrected for dust attenuation.  

Column 4: total (UV-based) SFR, corrected for dust attenuation using IR.

Column 5: SFR surface density, corrected for dust attenuation.

Column 6: HI gas surface density.

Column 7: HI reference.

Column 8: H$_{2}$ gas surface density.

Column 9: H$_{2}$ reference.

\subsection{Other properties\label{sec:otherprops}}
In order to investigate possible second-order correlations in the star formation law in Section~\ref{sec:secondorder}, we obtain measurements of various secondary parameters.

\begin{deluxetable*}{llccccccc}
\tablecolumns{9} 
\tablecaption{ Other properties of sample. \label{tab:otherprops}} 
\tablehead{ 
\colhead{N} & \colhead{NED ID} & \colhead{$\log M_{*}$} & \colhead{$M_{*}$ reference} & \colhead{$Z$} & \colhead{$Z$ reference} & \colhead{$C_{42}$} & \colhead{$\tau_{\textrm{dyn}}$} & \colhead{$v_{\textrm{rot}}$} \\
\colhead{} & \colhead{} & \colhead{$[M_{\odot}]$} & \colhead{} & \colhead{($12+\log[\textrm{O/H}]$)} & \colhead{} & \colhead{} & \colhead{($10^{8}$~yr)} & \colhead{(kpc/$10^{8}$~yr)} 
}
\startdata
1 & WLM & 7.39 & 1 & 7.83 & 2 & \ldots & 1.92 & 1.12 \\
2 & NGC 7817 & 10.55 & 3 & \ldots & \ldots & 2.53 & 1.90 & 21.97 \\
3 & NGC 0023 & \ldots & \ldots & 8.50 & 4 & \ldots & 1.75 & 26.88 \\
4 & UGC 00191 & 8.82 & 3 & \ldots & \ldots & 2.96 & \ldots & \ldots \\
5 & MESSIER 031 & \ldots & \ldots & \ldots & \ldots & \ldots & 3.18 & 22.48 \\
6 & IC 1574 & 7.74 & 1 & \ldots & \ldots & 2.40 & \ldots & \ldots \\
7 & NGC 0253 & 10.88 & 1 & 9.00 & 5 & \ldots & 3.36 & 21.46 \\
8 & UGCA 015 & 6.79 & 1 & \ldots & \ldots & 2.33 & \ldots & \ldots \\
9 & NGC 0278 & \ldots & \ldots & 8.47 & 4 & \ldots & 0.49 & 16.56 \\
10 & UGC 00634 & 9.03 & 3 & \ldots & \ldots & 2.90 & \ldots & \ldots
\enddata
\tablecomments{Only a portion of Table~\ref{tab:otherprops} is shown here; it is published in its entirety in the machine-readable format online.}
\tablereferences{ For stellar masses: (1) IR fluxes from LVL \citep{dale09}; (2) IR fluxes from SINGS \citep{dale07}; (3) S4G \citep{MunozMateos13, Querejeta15, Sheth10}. For metallicities: (1) \citet{Moustakas10}; (2) \citet{Berg12}; (3) \citet{Lee06}; (4) Calculated from spectra from \citet{Moustakas06} using \citetalias{PT05} calibration; (5) \citet{Cook14}. References for dynamical time and rotational velocity are largely the same as the HI and H$_{2}$ references listed in Table~\ref{tab:photometry}.}
\end{deluxetable*}

\subsubsection{Stellar mass}
We determine approximate stellar masses from mid-IR luminosities.
Several studies have found a nearly constant ratio of $M_{*}$ to 3.6~$\mu$m luminosity \citep[e.g.][]{Cook14,Eskew12}.
\citet{Eskew12} also use 4.5~$\mu$m luminosities; since observations from both wavelengths are available for the LVL and SINGS galaxies, we calculate stellar masses for these ($N = 130$) galaxies using the \citet{Eskew12} prescriptions:
\begin{equation*}
\frac{M_{*}}{M_{\odot}} = 10^{5.65}\left(\frac{F_{3.6}}{\textrm{Jy}}\right)^{2.85}\left(\frac{F_{4.5}}{\textrm{Jy}}\right)^{-1.85}\left(\frac{D}{0.05\textrm{Mpc}}\right)^{2}
\end{equation*}
For $N = 96$ additional galaxies, we obtain 3.6~$\mu$m-based stellar masses from the Spitzer Survey of Stellar Structure in Galaxies \citep[S4G;][]{Sheth10,MunozMateos13,Querejeta15}.
To check that the 4.5~$\mu$m and 3.6~$\mu$m calibration used for LVL and SINGS galaxies is consistent with the 3.6~$\mu$m calibration used in S4G, we compare the two calibrations for the LVL and SINGS galaxies where both fluxes are available. 
We find that the average difference between the two calibrations is $<0.005$~dex.

\subsubsection{Metallicity}
We compile $71$ integrated gas-phase metallicities\footnote{Unless otherwise noted, we refer to $Z=12+\log[\textrm{O/H}]$, or oxygen abundance relative to solar, as a proxy for total gas-phase metallicity.} from the following literature sources:
\citet{Berg12} and \citet{Lee06} compiled ``direct'' metallicities based on electron temperature $T_{e}$, while \citet{Moustakas10} used the \citet{PT05} (\citetalias{PT05}) ``strong-line'' calibration, which was empirically calibrated against direct metallicities. 

For an additional $64$ galaxies, we calculate metallicities from integrated optical spectra provided by \citet{Moustakas06}, using the \citetalias{PT05} calibration.
This calibration depends on the line ratio $R_{23} =  ([\textrm{O II}]\lambda\lambda3727,3729 + [\textrm{O III}]\lambda\lambda4959,5007)/\textrm{H}\beta$.
Since any relation between $R_{23}$ and $Z$ is double-valued, the line ratio $[\textrm{N II}]\lambda6854/\textrm{H}\alpha$ is used to determine which metallicity ``branch'' a galaxy occupies.

Finally, metallicities are obtained for a remaining $19$ galaxies from \citet{Cook14}.
Many of these are direct metallicities, but others were derived from inconsistent strong-line calibrations.
Due to this variation in calibration methods, \citet{Cook14} estimates potential systematic uncertainties of ${\sim}0.3$~dex for these abundances.

\begin{figure}
	\epsscale{1.2}
	\plotone{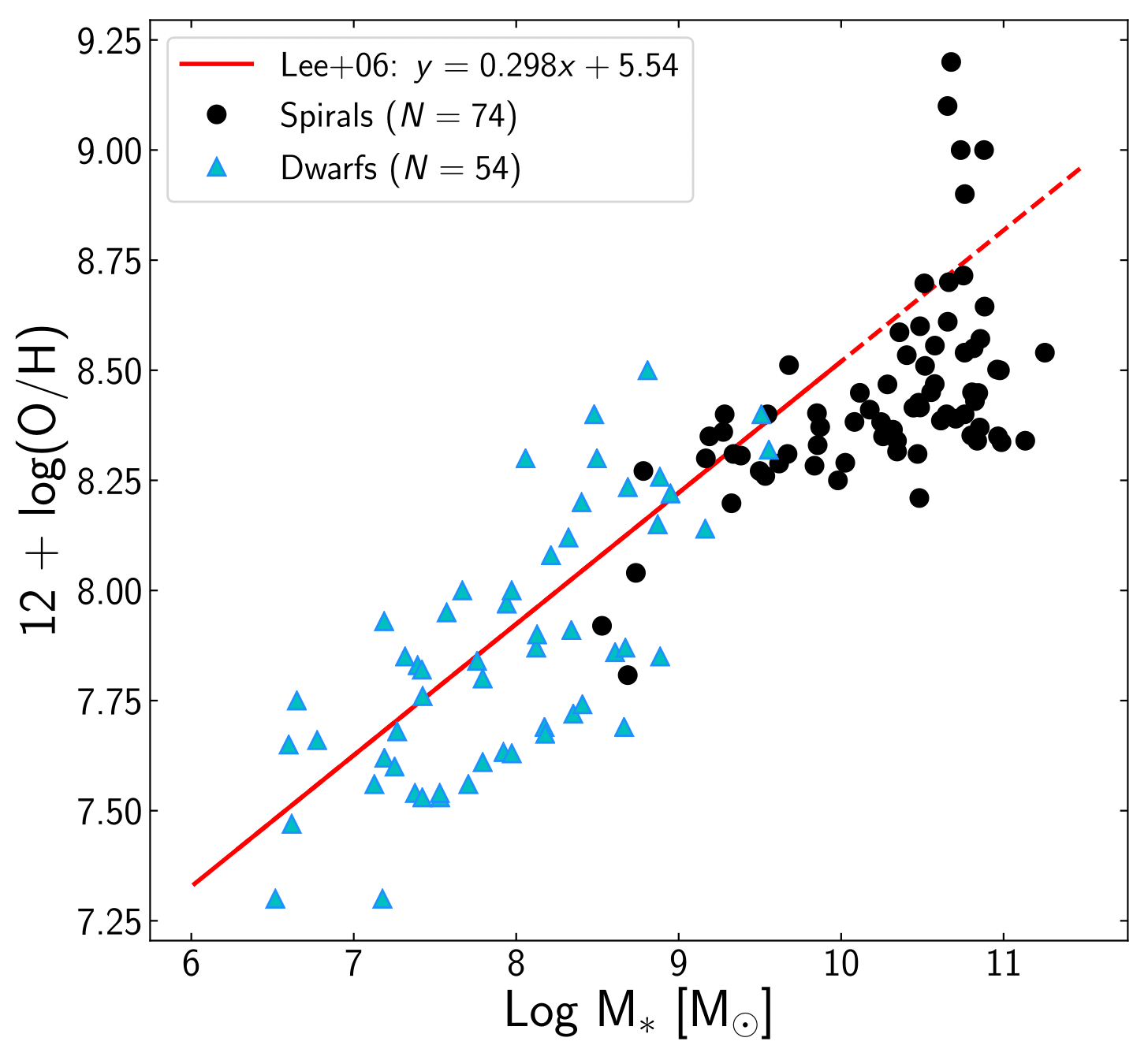}
	\caption{Mass-metallicity relation for our sample. The best-fit relation found for dwarf galaxies by \citet{Lee06} is plotted in red; the solid line illustrates the mass range over which this relation is valid, while the dashed line extrapolates this relation to the full mass range of our sample.\label{fig:MZR}}
\end{figure}

We characterize the stellar masses and gas-phase metallicities of the total sample by plotting the mass-metallicity relation in Figure~\ref{fig:MZR}. 
Our sample is consistent with the mass-metallicity relation determined for dwarf galaxies by \citet{Lee06}, which is a smooth extension of the well-studied mass-metallicity relation for star-forming galaxies \citep{Tremonti2004a}.

\subsubsection{Dynamical timescale\label{sec:tdyn}}
As described in Section~\ref{sec:intro}, the relationship between dynamical timescale ($\tau_{\textrm{dyn}}$) and SFR reflects a physical picture in which global perturbations convert a constant fraction of gas into stars \citep{Silk97}.
In this work, dynamical timescale is defined as the orbital timescale of the disk: $\tau_{\textrm{dyn}} = 2\pi R/v(R)$, where $R$ is the radius of the star-forming region, and $v$ is the rotational velocity.
Rotational velocities were compiled from the literature, with sources in the following order of preference: (1) full rotation curves measured in H$\alpha$, HI, and/or CO, either as published in the primary reference or measured by us from the data; (2) position-velocity relations in HI and/or CO; (3) 21\,cm HI linewidths, as compiled from NED, and corrected for turbulent broadening following the prescription of \citet{Tully85}.

All velocities were corrected for inclination, using values usually derived from fitting the velocity fields themselves, or otherwise from photometric measurements.  
We excluded galaxies less than 20\degr{} from face-on orientation, or systems with irregular or disturbed velocity fields.
The latter were especially problematic for some of the lowest-mass dwarf galaxies; as a result, the number of galaxies with $\tau_{\textrm{dyn}}$ ($N=163$) is considerably smaller than that for the gas surface density power law ($N=244$).
The rotation speed was measured at the radius of the star-forming disk, as given in Table~\ref{tab:photometry}, though for most galaxies the edge of the star-forming disk lies in the
flat part of the rotation curve.
This same radius was adopted to calculate the dynamical time.
Readers should be aware that the scaling radius for $\tau_{\textrm{dyn}}$ varies in the literature; the definition here is consistent with that used in \citetalias{Kennicutt98}.

\subsubsection{Concentration index}
To investigate how the star formation law might be affected by morphology, we consider the concentration index.
This index, often denoted $C_{42}$, is defined as
\begin{equation*}
C_{42} = 5\log\left(\frac{r_{80}}{r_{20}}\right),
\end{equation*}
where $r_{80}$ and $r_{20}$ are the radii containing 80\% and 20\% of a galaxy's light, respectively \citep{Kent85}.
This is a rough quantitative measure of galaxy morphology, since it is a proxy for the bulge-to-total luminosity ratio; elliptical (spiral) galaxies therefore have higher (lower) $C_{42}$ \citep{Shimasaku01}.
We obtain $C_{42}$ values for $N = 200$ galaxies from the S4G survey, which measured $C_{42}$ at 3.6~$\mu$m \citep{Sheth10}.

All of the secondary properties described in the previous sections are compiled in Table~\ref{tab:otherprops}:

Column 1: running index number.

Column 2: galaxy name preferred by the NASA/IPAC Extragalactic Database (NED).

Column 3: stellar mass.  

Column 4: stellar mass reference.

Column 5: gas-phase metallicity.

Column 6: gas-phase metallicity reference.

Column 7: concentration.

Column 8: dynamical timescale.

Column 9: inclination-corrected rotational velocity.

\subsubsection{Derived properties\label{sec:derivedprops}}
We note that additional important parameters can be derived from the ones listed in Tables~\ref{tab:UVdata}, \ref{tab:IRdata}, \ref{tab:photometry}, and \ref{tab:otherprops}.
These derived properties include: 
\begin{itemize} 
\item Specific star formation rate, SSFR = SFR/$M_{*}$
\item Molecular gas fraction, $f_{\textrm{mol}} = M_{\textrm{H}_{2}}/(M_{\textrm{HI}} + M_{\textrm{H}_{2}})$
\item Stellar mass surface density, $\Sigma_{*} = M_{*}/(\pi R^{2})$
\item Gas fraction, $f_{\textrm{gas}} = \frac{M_{\textrm{HI}} + M_{\textrm{H}_{2}}}{(M_{*} + M_{\textrm{HI}} + M_{\textrm{H}_{2}})}$
\item The ratio of obscured star formation to unobscured star formation, which is roughly approximated by $L_{\textrm{IR}}/L_{\textrm{UV}}$
\end{itemize}

\section{Star Formation Scaling Laws} 
\label{sec:results}

In this chapter, we update the global star formation law, consider alternative star formation laws, and discuss potential systematic uncertainties affecting our data.

\subsection{The revised star formation law for spiral galaxies}
\label{sec:spirals}
\begin{figure}
	\epsscale{1.2}
	\plotone{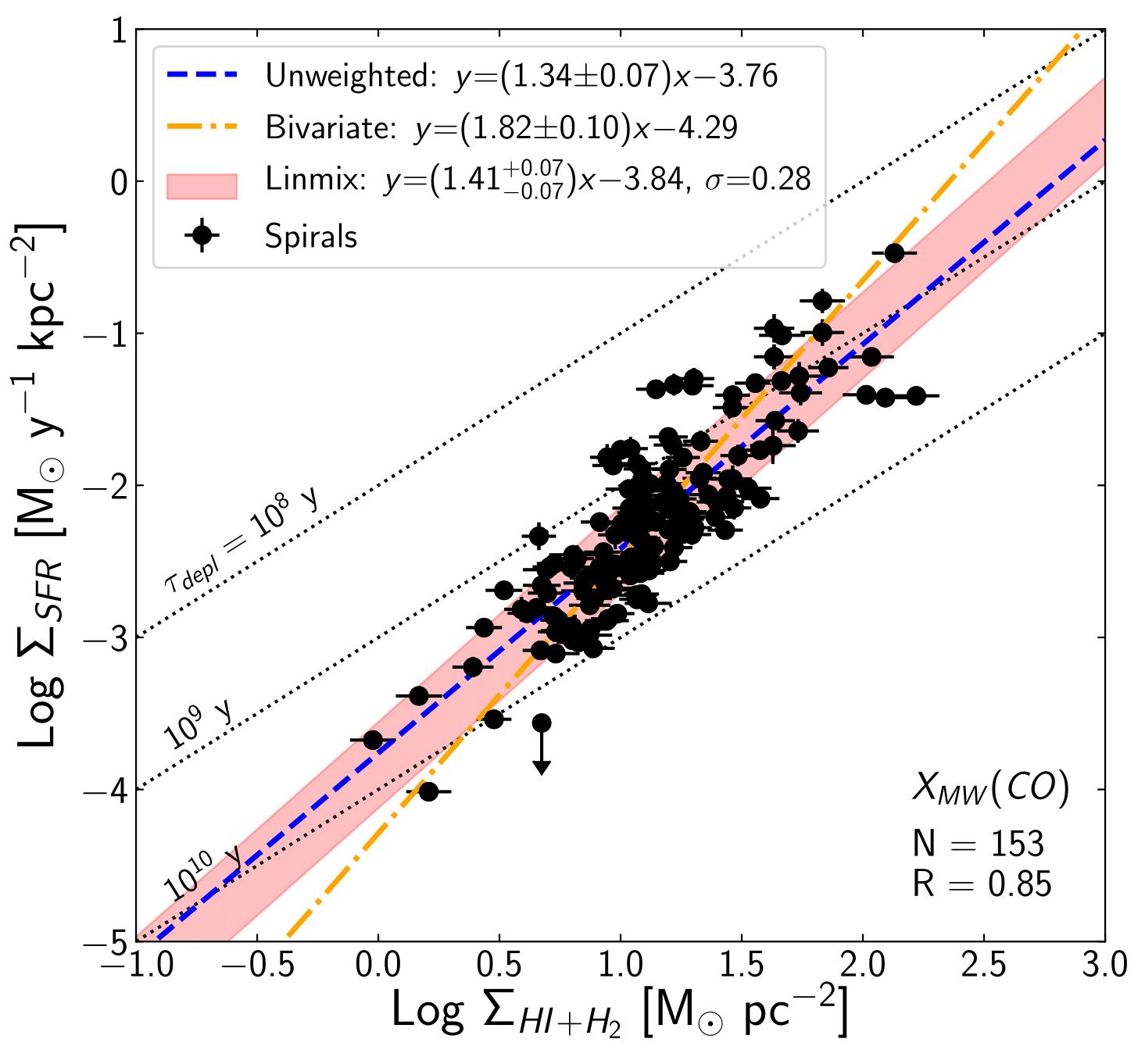}
	\caption{The global star formation law for spirals (black circles), using a constant Milky Way $X(\textrm{CO})$. Dashed lines represent constant depletion time $\tau_{\textrm{depl}}=\Sigma_{\textrm{gas}}/\Sigma_{\textrm{SFR}}$. The lines of best fit are derived using different methods as described in the text: \emph{unweighted} linear regression (blue dashed line), \emph{bivariate} linear regression (orange dot-dashed line), and an MCMC model using the \emph{linmix} algorithm (red shaded area marks the median of the posterior distributions for the linear slope, intercept, and intrinsic dispersion). Note that the correlation coefficient $R$ shown in the bottom right is the Pearson correlation coefficient. \label{fig:KSlawspirals}}
\end{figure}

We first repeat the analysis of \citetalias{Kennicutt98} and consider the relationship between $\Sigma_{\textrm{SFR}}$ and total gas surface density $\Sigma_{\textrm{gas}}$ using a constant Milky Way value of $X(\textrm{CO})$ (Equation~\ref{eq:XCO_MW}).
In this paper, as in many studies of the star formation law, we define $\Sigma_{\textrm{gas}}=\Sigma_{\textrm{HI}}+\Sigma_{\textrm{H}_{2}}$ and ignore contributions from helium and other species.
In Figure~\ref{fig:KSlawspirals}, we present the global star formation law for ``typical'' spiral galaxies (black solid points) and determine lines of best fit to this relation.
Although a thorough analysis of statistical methods to fit a line is beyond the scope of this paper \citep[for a more detailed discussion, see][]{Hogg10}, we briefly discuss the techniques used here. 

We first consider a naive \emph{unweighted} linear regression, which simply minimizes the mean squared errors of the $y$-residuals and weights all points equally, without accounting for any statistical uncertainties. 
This is clearly an inadequate model, since there are heteroscedastic measurement uncertainties in both the $x$ and $y$ directions.
However, since the errors in $\Sigma_{\textrm{SFR}}$ and the errors in $\Sigma_{\textrm{gas}}$ are of roughly the same order of magnitude, this method (dashed blue line in Figure~\ref{fig:KSlawspirals}) provides a first-order approximation to compare with other fits, yielding a slope of $1.34\pm 0.07$.
Note that all parameter uncertainties reported in this text are $1\sigma$ fitting errors.

We then consider measurement uncertainties in both $x$- and $y$-directions by computing a \emph{bivariate} fit using orthogonal distance regression.
Note that we consider only statistical measurement uncertainties here; we discuss systematic uncertainties in Section~\ref{sec:uncertainties}.
This method minimizes the orthogonal squared distance from the line to all points, which are weighted by the uncertainties in both the $x$- and $y$-directions. 
Bivariate regression was used by \citetalias{Kennicutt98} to determine power-law indices of $n=1.4\pm 0.15$ for spiral and starburst galaxies and $n=2.47\pm 0.39$ for spiral galaxies alone.
When applied to our updated sample of spiral galaxies (dot-dashed orange line in Figure~\ref{fig:KSlawspirals}), this regression yields an intermediate slope of $1.82\pm 0.10$.

However, it can be shown analytically that bivariate regression becomes biased and tends to overestimate the slope when there is intrinsic dispersion in the relation between the $x$ and $y$ quantities \citep{Akritas96, Carroll96}.
To more appropriately handle both $x$- and $y$-errors as well as intrinsic dispersion in the relation, we therefore use a hierarchical Bayesian model called \emph{linmix}, described by \citet{Kelly07}.\footnote{The linmix algorithm has been ported to a Python package by J. Meyers and is available on github at \url{https://github.com/jmeyers314/linmix}.}

The linmix method first assumes that a measured data point $(x,y)$ can be drawn from a two-dimensional Gaussian distribution $P_{1}$ with some ``true'' mean $(\xi,\eta)$ and covariance matrix determined from measurement uncertainties $\sigma_{x}$ and $\sigma_{y}$.
The ``true'' value of the dependent variable $\eta$ can in turn be drawn from a Gaussian distribution $P_{2}$ with mean $\beta\xi + \alpha$ and variance $\sigma^{2}$, where $\beta$ describes the slope of the line, $\alpha$ the $y$-intercept, and $\sigma^{2}$ the intrinsic dispersion in the $y$-direction.
Finally, the ``true'' value of the independent variable $\xi$ is assumed to be drawn from a weighted sum of $K$ Gaussian distributions $P_{3}$, since a large enough number of Gaussians can approximate any true distribution.\footnote{Following the procedure of \citet{Kelly07}, we use $K=2$ Gaussians to estimate posteriors. The addition of more Gaussians has a negligible effect on our results.}
The distributions $\{P_{1},P_{2},P_{3}\}$ are then convolved hierarchically to compute the full likelihood of obtaining the data $(x,y)$ given parameters $\{\beta,\alpha,\sigma^{2}\}$.
Assuming uniform prior distributions for the parameters $\{\beta,\alpha,\sigma^{2}\}$, a Markov chain Monte Carlo (MCMC) algorithm is used to sample from the posterior distributions until convergence is reached.
We note that for all linmix fits presented in this paper, the marginalized posterior distributions for $\{\beta,\alpha,\sigma^{2}\}$ are roughly Gaussian.

This linmix model (red shaded area in Figure~\ref{fig:KSlawspirals}) yields a median slope of $1.41^{+0.07}_{-0.07}$, where the parameter uncertainties given are the 16th and 84th percentiles of the marginalized distribution.
This slope is much shallower than the bivariate \citetalias{Kennicutt98} result for the spiral galaxies alone ($n=2.47\pm 0.39$).
It is, however, remarkably consistent with the oft-cited $n\sim 1.4$ law for both spiral and starburst galaxies.
We take this result to be the fiducial star formation law for spiral galaxies:
\begin{equation}
\log \Sigma_{\textrm{SFR}} = \left(1.41^{+0.07}_{-0.07}\right)\log \Sigma_{\textrm{gas}} - 3.84^{+0.08}_{-0.09}.
\label{eq:KSlawspirals}
\end{equation}
Furthermore, the linmix method estimates an intrinsic dispersion in the $y$-direction of $\sigma = 0.28^{+0.02}_{-0.02}$~dex in the y-direction, which is larger than the typical measurement uncertainty in SFR ($\sigma_{\textrm{SFR}}\lesssim 0.10$~dex).

The linmix model also estimates the intercept of the star formation law to be $-3.84^{+0.08}_{-0.09}$, which corresponds to a coefficient $A=1.5^{+0.3}_{-0.3}\times 10^{-4}$ in Equation~\ref{eq:SchmidtLaw}.
This is slightly smaller than the value $A=2.5\pm 0.7$ measured in \citetalias{Kennicutt98}.
This discrepancy may result from a number of factors.
In particular, the composite FUV and 24\micron{} calibrations used in this work yield SFRs ${\sim}0.2$~dex lower than those in \citetalias{Kennicutt98}, primarily due to differences in the assumed IMF and updated stellar population models \citep{Kennicutt12rev}.
The individual 24\micron{} dust corrections used to compute SFRs may also play a role; we correct the UV luminosities for internal extinction by a factor of ${\sim}2.4$ on average, compared to the factor of $2.8$ assumed by \citetalias{Kennicutt98}.
Some combination of these explanations likely produces the discrepancy in measured star formation law intercepts.

We note that upper limits are excluded when applying all three fitting methods.
In our sample of spiral galaxies, there is only a single point with an upper limit in $\Sigma_{\textrm{SFR}}$ (see Figure~\ref{fig:KSlawspirals}), so we do not expect its exclusion to significantly affect the fit.
For completeness, we check this using the linmix method, which is capable of handling upper limits in the $y$-direction \citep{Kelly07}.
We find that including this upper limit does not significantly affect the model parameters.

\begin{figure*}
	\epsscale{1.15}
	\plottwo{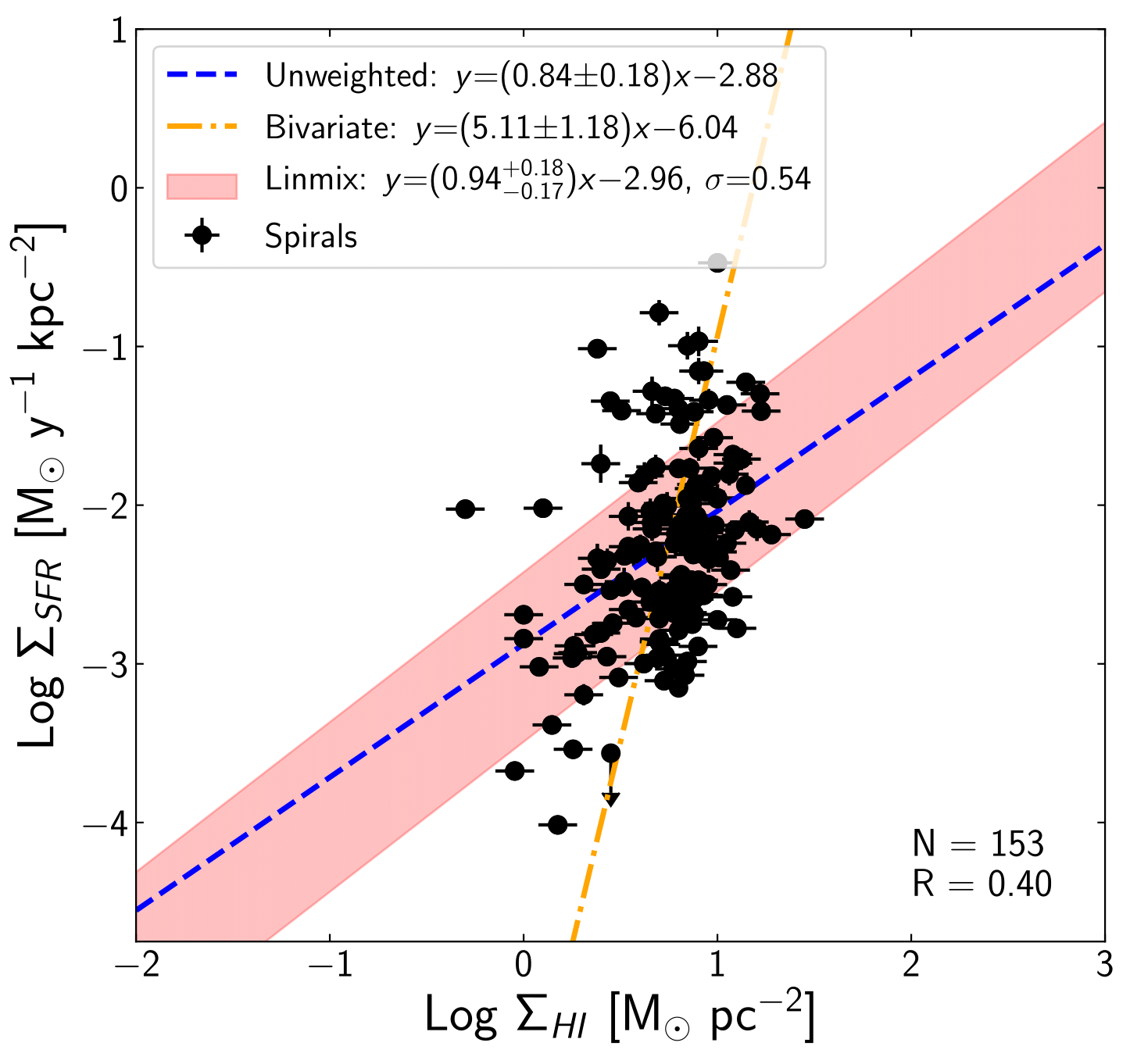}{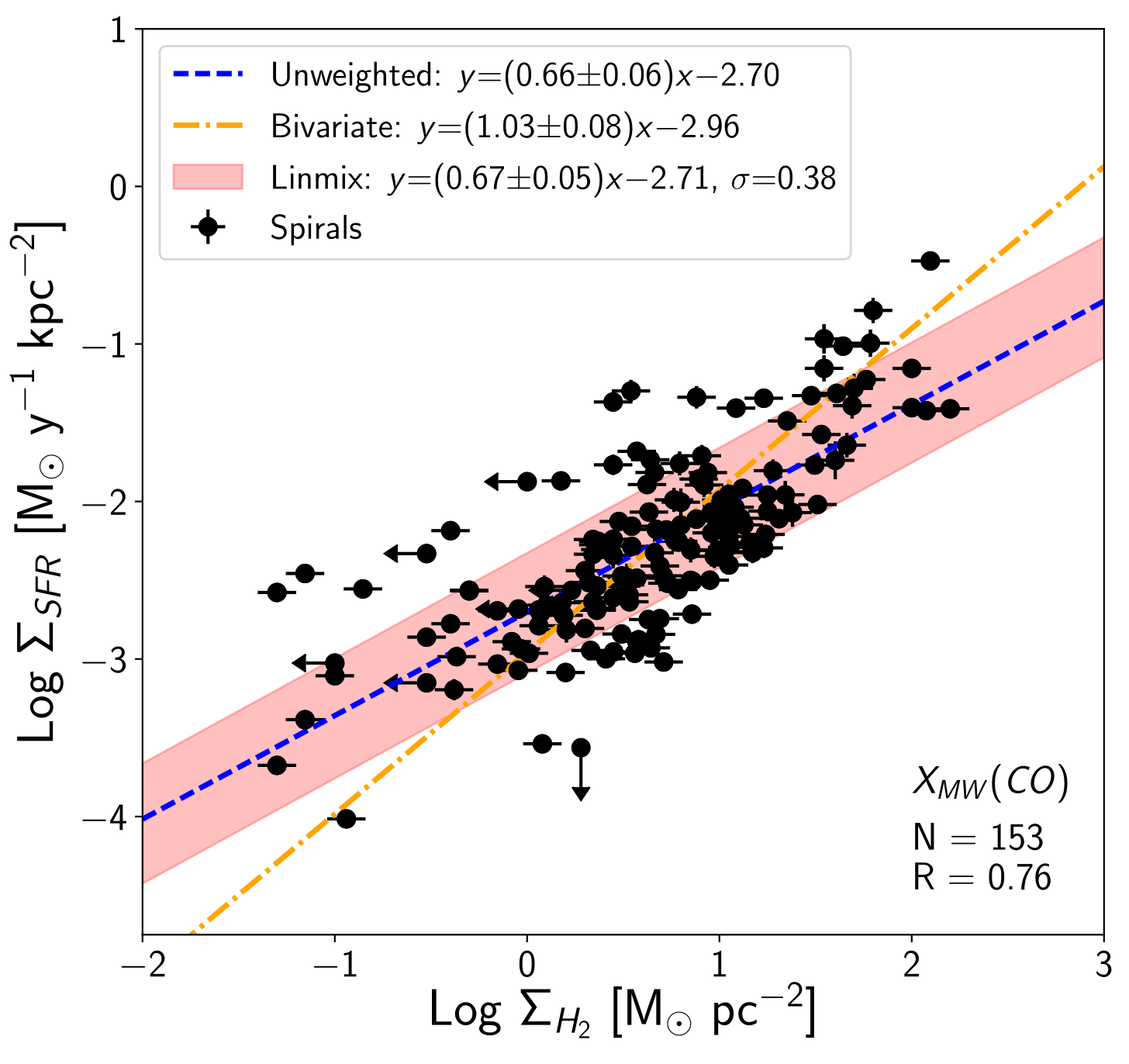}
	\caption{The relationships between SFR surface density and atomic gas surface density (left) and between SFR surface density and molecular gas surface density (right) for spiral galaxies (black circles). Downward arrows represent upper limits on $\Sigma_{\textrm{SFR}}$. Linear fits are marked by lines and shaded regions as described in Figure~\ref{fig:KSlawspirals}.\label{fig:KSatomicmol}}
\end{figure*}

\subsection{Separate atomic and molecular gas components in spiral galaxies}
\label{sec:molatomSFlaws}

As in \citetalias{Kennicutt98}, we now separately consider the atomic and molecular hydrogen gas components.
Star formation is thought to occur in molecular clouds; indeed, results from spatially-resolved studies of the star formation law have found moderate correlation between $\Sigma_{\textrm{SFR}}$ and $\Sigma_{\textrm{H}_{2}}$ but little to no correlation between $\Sigma_{SFR}$ and $\Sigma_{\textrm{HI}}$ \citep{Bigiel08, Kennicutt07, 101, 225}. 
Several of these studies have proposed that star formation is a two-step process: a baryonic reservoir of atomic gas is converted to molecular gas, and dense clumps of molecular gas are then converted into stars.

On the global scale, \citetalias{Kennicutt98} found that the correlation between SFR density and atomic gas density was almost as strongly correlated as the $\Sigma_{\textrm{SFR}}$-$\Sigma_{\textrm{HI}+\textrm{H}_{2}}$ relation, with a Pearson correlation coefficient\footnote{In this analysis, all correlation coefficients are Pearson product-moment correlation coefficients.} of $R = 0.66$ (compared to $R = 0.68$ for the total gas relation). 
The correlation between SFR density and molecular gas density was much weaker, and it was suggested that this was perhaps in part due to variations in the CO-to-H$_{2}$ conversion factor $X(\textrm{CO})$. 
We investigate the tension between these conflicting results with our updated sample.

The left panel of Figure~\ref{fig:KSatomicmol} illustrates the relationship between $\Sigma_{\textrm{SFR}}$ and $\Sigma_{\textrm{HI}}$ for spiral galaxies.
The correlation between the SFR and atomic gas surface density is weaker than that found by \citetalias{Kennicutt98}, with a correlation coefficient $R=0.40$ compared to the \citetalias{Kennicutt98} value of $R=0.66$.
The large difference between the slopes estimated using various regression methods ($n = 5.10\pm 1.20$, $n=0.94^{+0.18}_{-0.17}$ and $n = 0.84\pm0.18$ for bivariate, linmix, and unweighted regressions, respectively) also suggests a weak $\Sigma_{\textrm{SFR}}$-$\Sigma_{\textrm{HI}}$ correlation, as does the large intrinsic dispersion $\sigma=0.54^{+0.03}_{-0.03}$ estimated by linmix. 
These inconsistencies suggest that a simple power law fit is a poor representation of the observed relation. 
Instead we suspect that the nearly vertical relation is strongly influenced by the conversion of atomic hydrogen to molecular gas above a surface density of ${\sim}10~\textrm{M}_{\odot}~\textrm{pc}^{-2}$ (column density of ${\sim}10^{21}~\textrm{cm}^{-2}$), as discussed previously for spatially-resolved studies by \citet{Kennicutt07} and \citet{Bigiel08}.

We now consider the $\Sigma_{\textrm{SFR}}$-$\Sigma_{\textrm{H}_{2}}$ relation for spiral galaxies in the right panel of Figure~\ref{fig:KSatomicmol}.
As noted in Section~\ref{sec:gasdata}, the molecular gas surface density depends on the assumed $X(\textrm{CO})$ conversion factor from CO luminosity to H$_{2}$ gas mass; we initially assume a constant $X(\textrm{CO})$ that is roughly reliable for non-starbursting spiral galaxies (Equation~\ref{eq:XCO_MW}).
The correlation between $\Sigma_{\textrm{SFR}}$ and molecular gas density $\Sigma_{\textrm{H}_{2}}$ is stronger than the star formation law based on atomic gas density, with a correlation coefficient of $R = 0.79$. 
Although the various regression methods find inconsistent slopes, again suggesting some amount of intrinsic scatter in the relation, all methods yield approximately linear slopes $n\sim 1$.

Both the weak $\Sigma_{\textrm{SFR}}-\Sigma_{\textrm{HI}}$ correlation and tighter, roughly linear $\Sigma_{\textrm{SFR}}-\Sigma_{\textrm{H}_{2}}$ correlation are consistent with the results found by spatially-resolved studies \citep[e.g.,][]{Bigiel08}.
These are the opposite of the global results reported by \citetalias{Kennicutt98}, suggesting that the \citetalias{Kennicutt98} results were strongly influenced by a smaller sample size and/or a narrower parameter space.
However, both the atomic gas and the molecular gas star formation laws have weaker correlations than the total gas star formation law in Figure~\ref{fig:KSlawspirals} ($R=0.85$); the linmix method also finds that the atomic gas and molecular gas star formation laws have larger intrinsic dispersions by $\geq 0.1$~dex.
This appears to confirm the \citetalias{Kennicutt98} finding that $\Sigma_{\textrm{HI}+\textrm{H}_{2}}$ is a strong predictor of SFR surface density.

\subsection{Molecular gas conversion factors}
\label{sec:xco}
Before extending the star formation law to include low-$\Sigma_{\textrm{SFR}}$ dwarf galaxies, we must consider the effect of varying $X(\textrm{CO})$.
$X(\textrm{CO})$ is likely dependent on environmental factors; in particular, metallicity is correlated with the presence of dust, which can help shield CO from photodissociation.
As a result, in low-mass and low-metallicity environments such as dwarf galaxies, CO will under-predict the true amount of $\textrm{H}_{2}$, and $X(\textrm{CO})$ should be higher than the Milky Way value \citep[see][and references therein]{Bolatto13}.
We therefore consider the effect of using metallicity-dependent prescriptions of $X(\textrm{CO})$.

First, we consider the prescription recommended by \citet{Bolatto13}, who used a simple analytical model to approximate a correction factor $f$ such that $X(\textrm{CO})$ = $fX_{\textrm{MW}}(\textrm{CO})$:
\begin{equation}
f=0.67\exp\left(\frac{0.4}{Z'\Sigma_{\textrm{100,GMC}}}\right).
\label{eq:XCO_Bolatto}
\end{equation}
Here, $Z' = Z/Z_{\odot}$ is metallicity relative to solar \citep[note that 12+log(O$_{\odot}$/H$_{\odot}$) = 8.69;][]{Asplund09}, and $\Sigma_{\textrm{100,GMC}} = 1.0$ is the assumed characteristic surface density of molecular clouds in units of 100~M$_{\odot}$~pc$^{-2}$. This factor accounts for the H$_{2}$ mass in outer regions of clouds, where CO is more likely to be photodissociated.

\citet{Glover10} developed a more extreme $X(\textrm{CO})$ prescription by simulating a more complex dynamical model for individual giant molecular clouds.
We use the \citet{Bolatto13} adaptation of Equation~16 in \citet{Glover10}, which assumes that mean extinction (i.e., dust abundance) scales linearly with metal abundance.
This yields the following correction factor $f$:
\[ f =
  \begin{cases}
    1 & \text{for $Z'\overline{A}_{V,MW} > 3.5$} \\
    (\frac{Z'\overline{A}_{V,MW}}{3.5})^{-3.5} & \text{for $Z'\overline{A}_{V,MW} < 3.5$},
  \end{cases}
\]
where $Z'$ is again metallicity relative to solar, and $\overline{A}_{V,MW}=5$ is the assumed mean extinction through a molecular cloud with surface density $\Sigma^{100}_{GMC} = 1.0$.

\begin{figure}
	\epsscale{1}
	\plotone{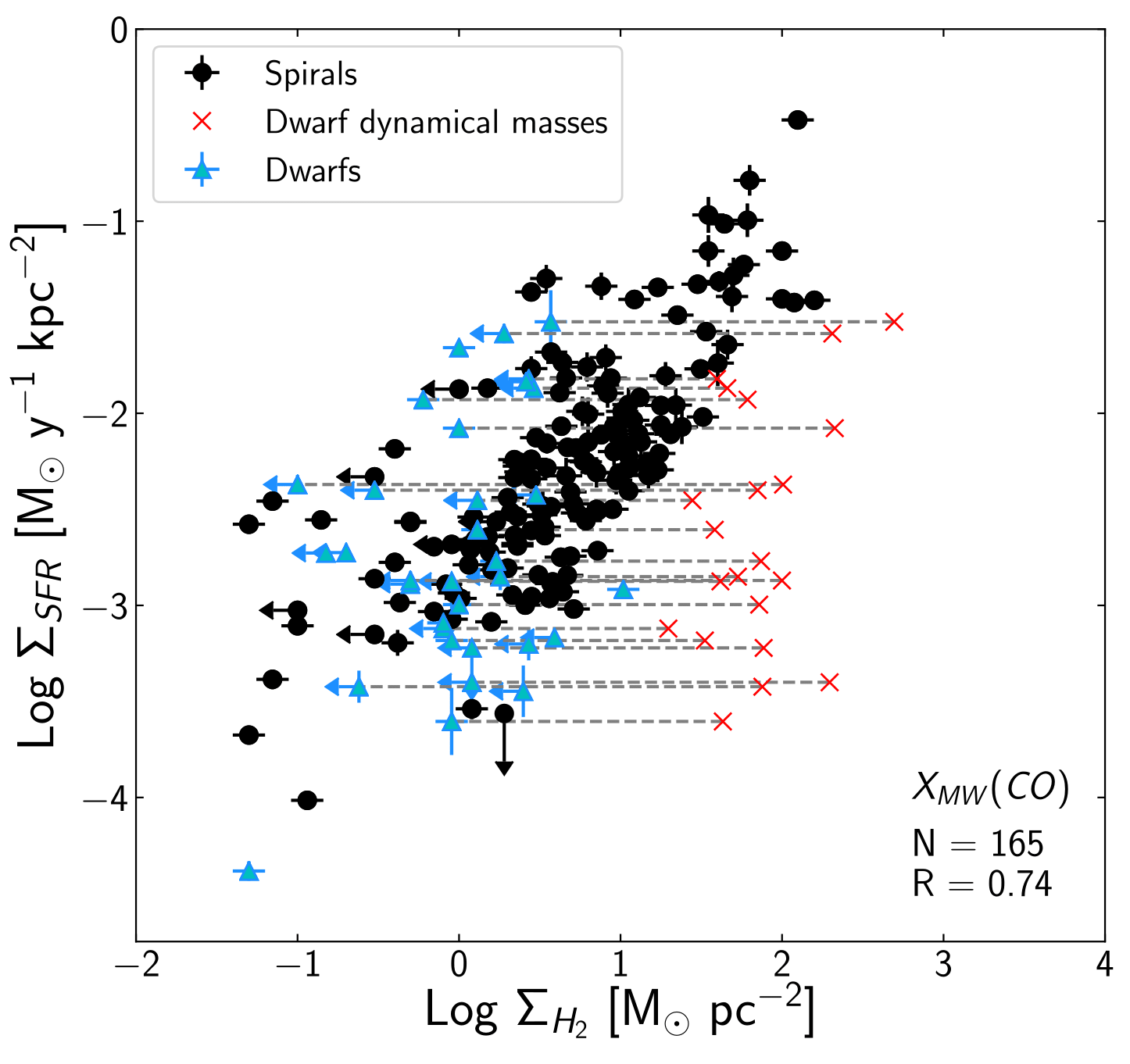}
	\plotone{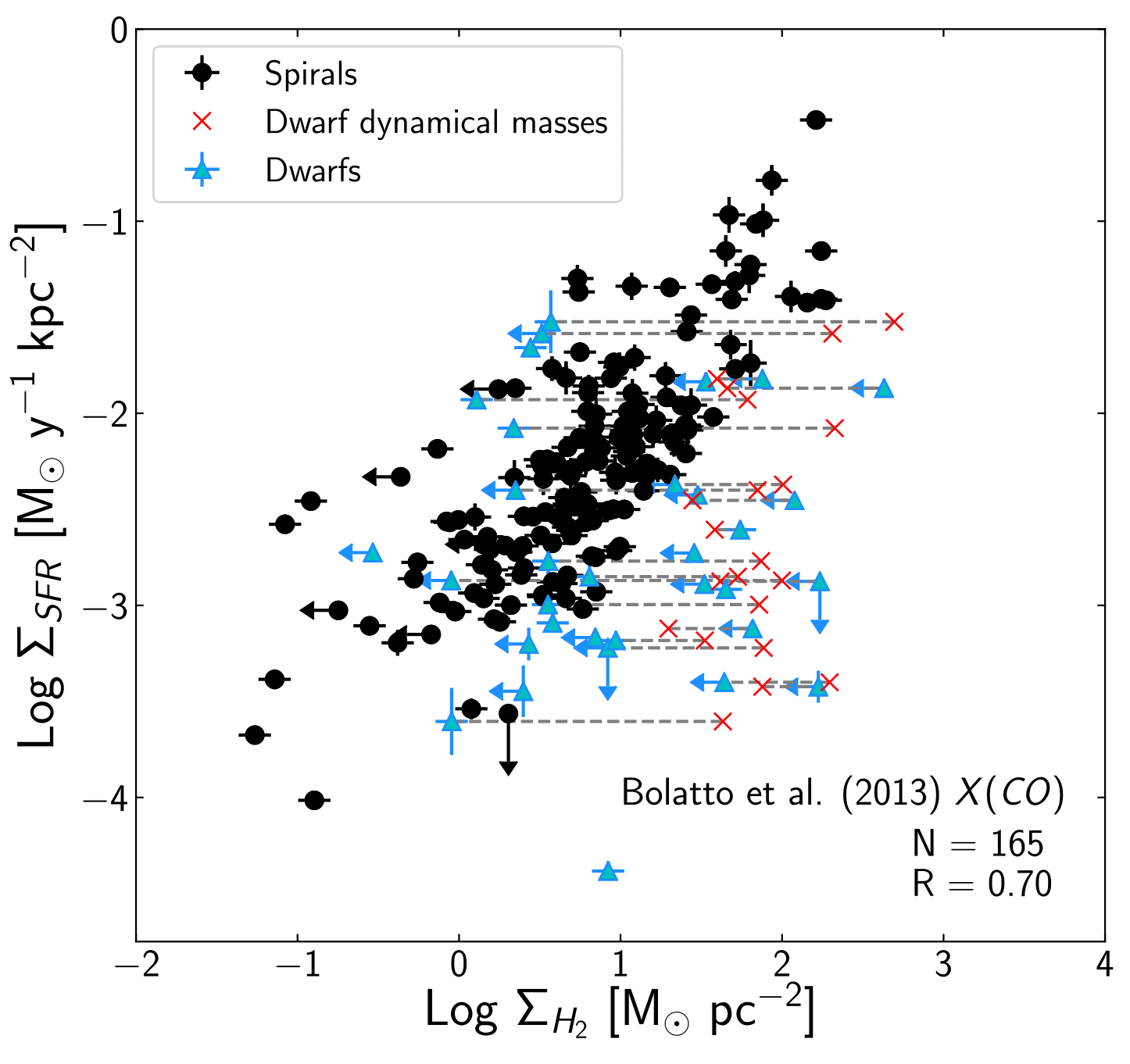}
	\plotone{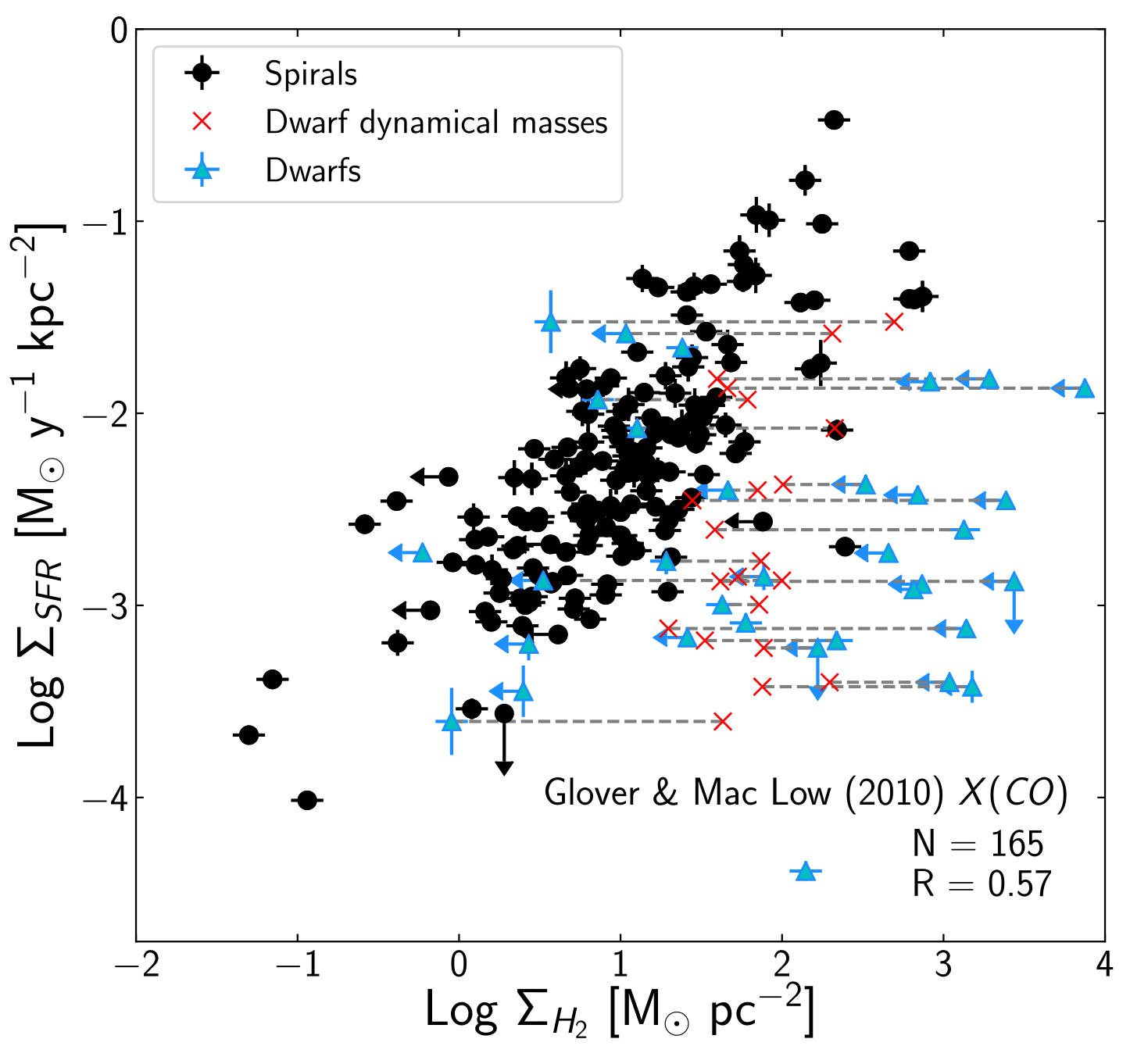}
	\caption{The effect of using various $X(\textrm{CO})$ prescriptions (described in the text) on the relationship between SFR surface density and molecular gas surface density for spirals (black circles) and dwarfs (cyan triangles). For each dwarf galaxy with an available dynamical mass, the dynamical mass surface density $\Sigma_{\textrm{dyn}}$ is plotted as a connected red X. Arrows represent upper limits on either $\Sigma_{\textrm{H}_{2}}$ (downward) or $\Sigma_{\textrm{SFR}}$ (leftward). \label{fig:KSmolecular}}
\end{figure}

We note that even spiral galaxies are affected by changing $X(\textrm{CO})$.
Although the relationship between star formation rate and \emph{molecular} gas surface density for spirals does not change significantly, using metallicity-dependent $X(\textrm{CO})$ prescriptions rather than a constant Milky Way value can decrease the slope of the \emph{total} gas star formation laws by ${\sim 0.12}$~dex, from $1.41$ to $1.29$. 

This effect can become particularly extreme for dwarf galaxies.
In Figure~\ref{fig:KSmolecular} we plot the relationship between $\Sigma_{\textrm{SFR}}$ and $\Sigma_{\textrm{H}_{2}}$ for these different $X(\textrm{CO})$ calibrations.
To check how realistic these prescriptions are, we compare the molecular gas surface densities $\Sigma_{\textrm{H}_{2}}$ with dynamical mass surface densities $\Sigma_{\textrm{dyn}}$, which are computed from dynamical mass $M_{\textrm{dyn}}=v^{2}R/G$.
Here $v$ is the circular velocity at the star formation radius $R$, computed from 21cm lines; see Section~\ref{sec:tdyn}.
For dwarf galaxies, $\Sigma_{\textrm{dyn}}$ are plotted as red Xs in Figure~\ref{fig:KSmolecular}.

Figure~\ref{fig:KSmolecular} shows that applying either the \citet{Bolatto13} or the \citet{Glover10} metallicity-dependent $X(\textrm{CO})$ prescription shifts the low-metallicity galaxies---which are mostly dwarf galaxies---to higher $\Sigma_{\textrm{H}_{2}}$.
Many of the shifted galaxies only have upper limits on $\Sigma_{\textrm{H}_{2}}$, so increasing these limits does not constrain the star formation law.
However, the \citet{Glover10} $X(\textrm{CO})$ formula has particularly extreme effects, shifting several dwarf galaxies without upper limits on $\Sigma_{\textrm{H}_{2}}$ to molecular gas surface densities greater than the dynamical mass surface densities. 
This unrealistic result is likely not because of any intrinsic flaw in the \citet{Glover10} prescription, but rather because their result is based on individual giant molecular clouds and is not applicable to entire galaxies.
As with the star formation rate calibrations (Section~\ref{sec:SFRcalibs}), we emphasize the importance of exercising caution when applying small-scale prescriptions to galaxy-scale problems.
We therefore use the \citet{Bolatto13} $X(\textrm{CO})$ prescription in all following analyses.

\begin{figure*}
	\epsscale{1.1}
	\plottwo{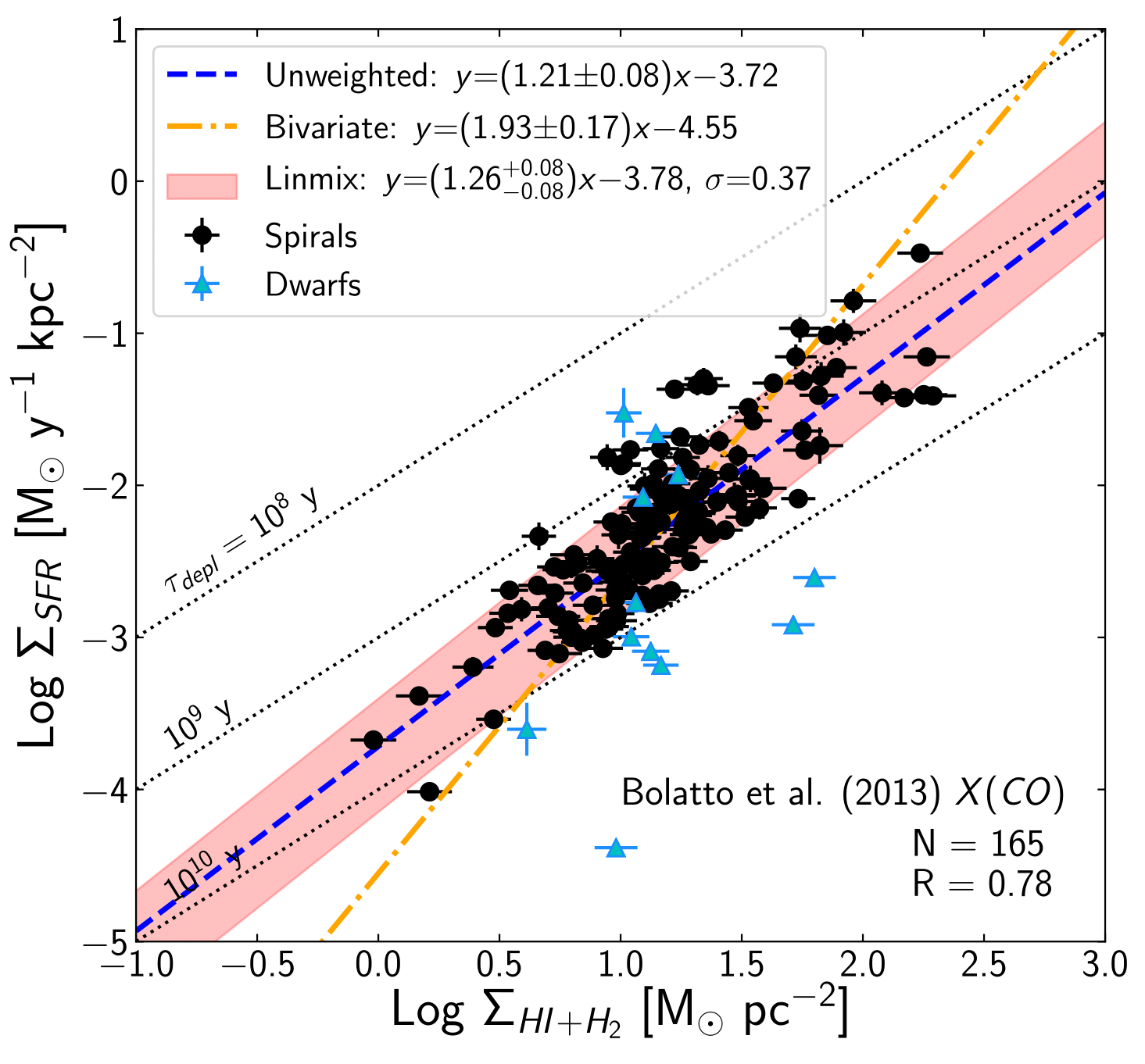}{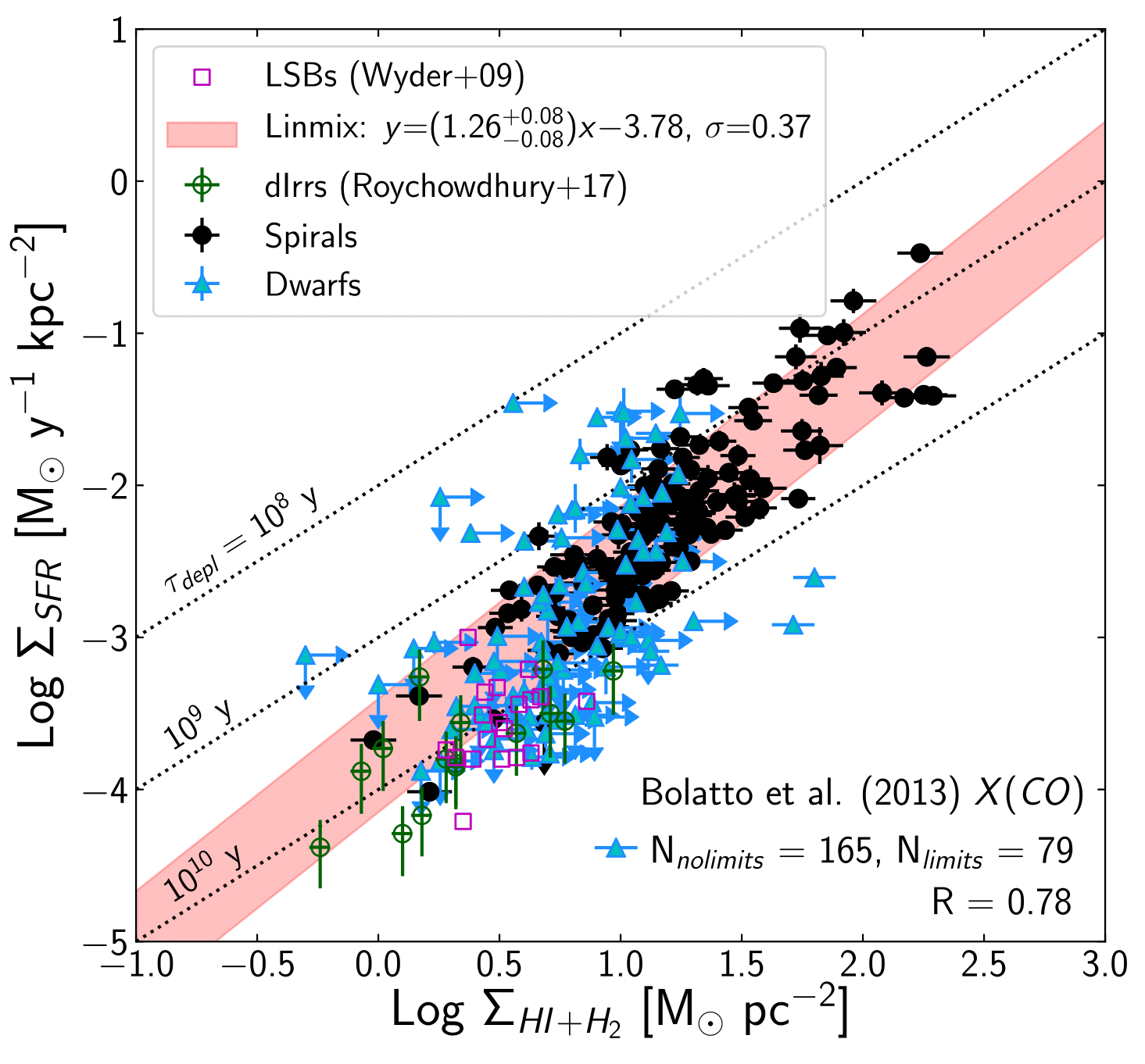}
	\caption{Same as Figure~\ref{fig:KSlawspirals}, but including dwarf galaxies (cyan triangles). The left panel only includes galaxies with CO and SFR detections, as well as unweighted (blue dashed line) and bivariate (yellow dot-dashed line) fits. The red shaded area marks the median fit and intrinsic dispersion determined using the linmix MCMC method. The right panel has the same linmix fit, but also plots galaxies with only HI measurements (rightward arrows) or upper limits on SFR measurements (downward arrows); the numbers of galaxies with and without these limits are given by $N_{\textrm{limits}}$ and $N_{\textrm{nolimits}}$, respectively. For comparison, low-surface brightness galaxies from \citet{Wyder09} (purple unfilled triangles) and dwarf irregular galaxies from \citet{Roychowdhury17} (green unfilled circles) are also shown.\label{fig:KSlawtotal}}
\end{figure*}

\subsection{The star formation law for both spiral and dwarf galaxies}
\label{sec:spiraldwarfs}
Having determined which $X(\textrm{CO})$ prescription is the least unrealistic for our sample, we can now extend the total gas star formation law to include low-$\Sigma_{\textrm{SFR}}$ dwarf galaxies (cyan triangles).
In the left panel of Figure~\ref{fig:KSlawtotal}, we plot all galaxies with HI, CO, and SFR measurements, showing a clear correlation between $\Sigma_{\textrm{SFR}}$ and $\Sigma_{\textrm{gas}}$ (correlation coefficient $R=0.81$).

Various linear regression methods produce a larger range of star formation indices.
The unweighted method yields a slope of $n=1.21\pm 0.08$, the bivariate method yields a slope of $n=1.94\pm 0.17$, and the linmix model estimates a slope $n=1.26^{+0.08}_{-0.08}$.
As described in Section~\ref{sec:spirals}, the linmix regression method is more appropriate in fitting relations with some intrinsic scatter \citep[e.g.,][]{Hogg10}.
We therefore take the linmix result to be the fiducial star formation law for the combined sample of spirals and dwarf galaxies:
\begin{equation}
\log \Sigma_{\textrm{SFR}} = \left(1.26^{+0.08}_{-0.08}\right)\log \Sigma_{\textrm{gas}} - 3.78^{+0.10}_{-0.10}.
\label{eq:KSlawfinal}
\end{equation}
The linmix model also estimates an intrinsic dispersion of $\sigma = 0.37^{+0.02}_{-0.02}$~dex, which we discuss in further detail in Section~\ref{sec:secondorder}.

For visualization purposes, in the right panel of Figure~\ref{fig:KSlawtotal} we plot galaxies with measurement limits: galaxies with upper limits on SFR, or galaxies with upper limits or non-detections of H$_{2}$ (and therefore lower limits on the total $\Sigma_{\textrm{gas}}$).
We note that in our analyses, we exclude these points when determining the correlation coefficient and linear fit \footnote{The linmix method is capable of handling upper limits in the $y$-direction; however, as in Section~\ref{sec:spirals}, only one of our points has \emph{only} an upper limit in the $y$-direction. All other points with upper limits in the $y$-direction also have lower limits in the $x$-direction. Properly treating such points is a complex problem beyond the scope of this statistical analysis, particularly since the lower $x$-limits are primarily dominated by systematic uncertainties in $X(\textrm{CO})$ anyway (see Section~\ref{sec:uncertainties} for further discussion of this systematic effect).}.
We instead plot the linmix linear fits from Figure~\ref{fig:KSlawtotal} to compare these measurement limits to the derived slope of $n=1.26^{+0.08}_{-0.08}$. 
We also plot low-surface brightness galaxies (LSBs; purple triangles) from \citet{Wyder09} and dwarf irregular galaxies (dIrrs; green diamonds) from \citet{Roychowdhury17} for comparison.

Figure~\ref{fig:KSlawtotal} indicates that dwarf galaxies tend to increase the intrinsic scatter in the star formation law.
Indeed, the linmix regression method suggests that $\sigma$ (the intrinsic dispersion in the $y$-direction) increases by $\sim0.1$~dex when dwarf galaxies with reliable measurements are included in the fit.
The inclusion of dwarf galaxies also decreases the slope in the star formation law from $n\sim1.41$ (for spiral galaxies alone) to $n\sim1.26$ (for both spirals and dwarfs).
Furthermore, Figure~\ref{fig:KSlawtotal} shows that the majority of dwarf galaxies---including LSBs, dwarf irregular galaxies, and dwarf galaxies with lower limits on $\Sigma_{\textrm{gas}}$---lie at lower $\Sigma_{\textrm{SFR}}$ than the star formation law defined by spiral galaxies.
Those dwarf galaxies that lie above the main locus of spiral galaxies tend to have non-detections of CO (rightward arrows); a ``dark'' (i.e., not traced by CO) molecular component could shift these galaxies to higher $\Sigma_{\textrm{gas}}$, into agreement with the other dwarf galaxies.
Note that the choice of $X(\textrm{CO})$ does not affect this qualitative result, because adding a correction for undetected molecular gas only drives these galaxies towards higher $\Sigma_{\textrm{gas}}$ (without changing $\Sigma_{\textrm{SFR}}$), and thus further below the star formation law for spiral galaxies.

This may be consistent with a low-density threshold in the star formation law. 
Such a threshold---below which the star formation law steepens or even breaks down---has been well-measured on spatially-resolved scales within galaxies \citep[e.g.,][]{Skillman87,Kennicutt89,Bigiel08}. 
As shown in Figure~\ref{fig:KSlawtotal}, this threshold does appear to exist on integrated scales, although it is much less distinct than that observed on spatially-resolved scales.
We discuss potential interpretations of this threshold in Section~\ref{sec:interpretations}.

\subsection{Alternative star formation laws}
\label{sec:alternativeSFlaws}

The original Schmidt law is not the only star formation scaling law. 
Equation~\ref{eq:SELaw} describes an alternative version of the star formation law, in which the dynamical timescale $\tau_{\textrm{dyn}}$ is taken to be the characteristic time of star formation \citep{Silk97, Elmegreen97}.
In this relationship, hereafter referred to as the Silk-Elmegreen relation, star formation
efficiency $\epsilon = \Sigma_{\textrm{SFR}}/\Sigma_{\textrm{gas}}$ depends linearly on $\tau_{\textrm{dyn}}$ rather than on local gas density $\Sigma_{\textrm{gas}}$.

\begin{figure}
	\epsscale{1.2}
	\plotone{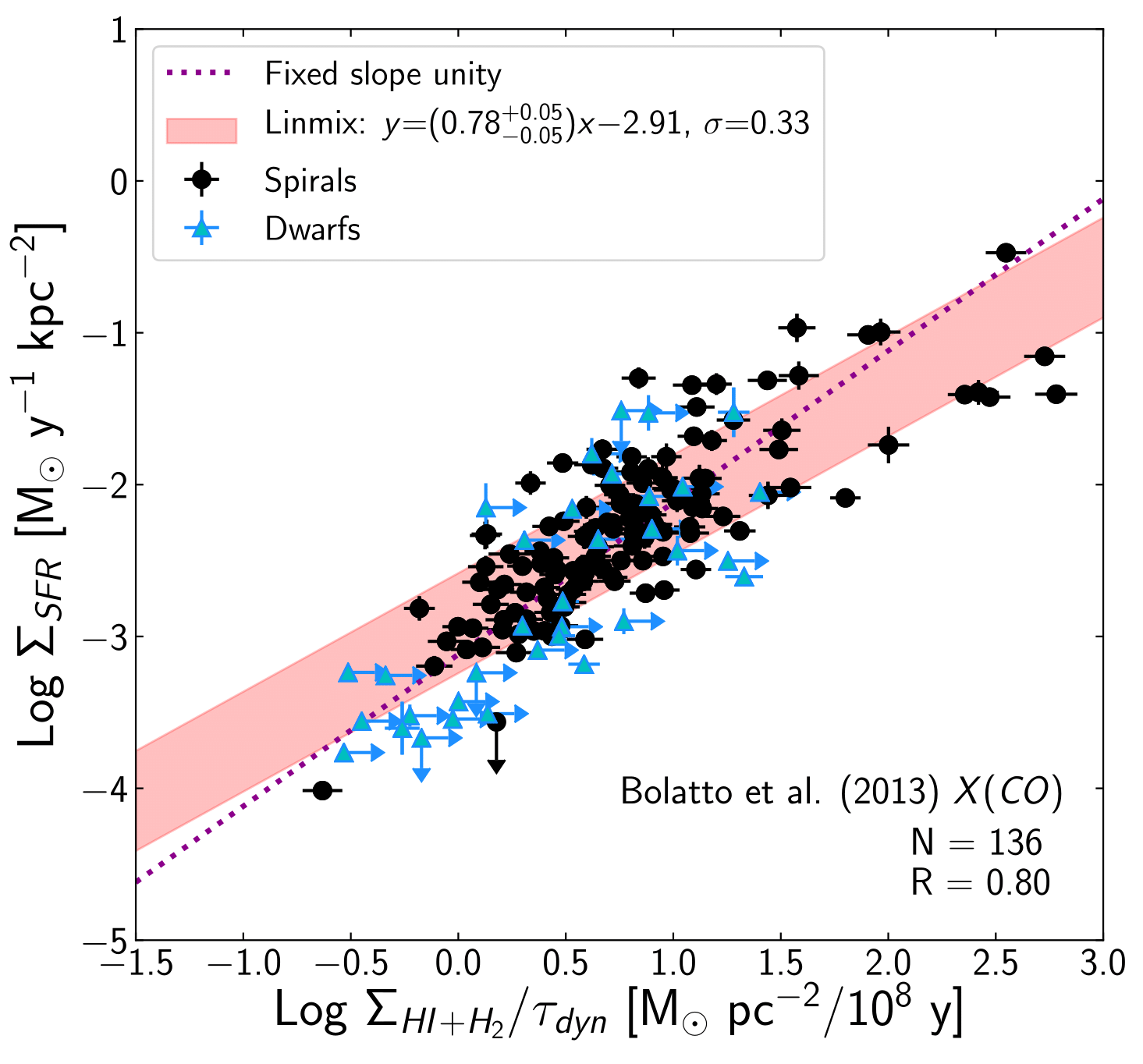}
	\caption{The Silk-Elmegreen relation for spirals (black circles) and dwarfs (cyan triangles), including galaxies with only HI measurements (rightward arrows) and upper limits on SFR measurements (downward arrows). The metallicity-dependent $X(\textrm{CO})$ prescription from \citet{Bolatto13} is used. Red shaded area marks the median fit and intrinsic dispersion computed using the linmix MCMC. The dotted purple line indicates a line of slope unity. \label{fig:SElaw}}
\end{figure}

We plot the Silk-Elmegreen relation for both our spiral and dwarf galaxies in Figure~\ref{fig:SElaw}.
As in \citetalias{Kennicutt98}, the local dynamical timescale $\tau_{\textrm{dyn}}$ is taken to be the orbital period at the star-forming radius (Section~\ref{sec:tdyn}). 
The slope of this relation estimated from the linmix algorithm is sub-linear ($n=0.78\pm0.05$), inconsistent with the predicted linear slope (dotted purple line). 
The shallower slope appears to be driven by spiral galaxies with high SFR surface densities that have higher $\Sigma_{\textrm{gas}}/\tau_{\textrm{dyn}}$ than expected, suggesting that the fraction of gas converted into stars is not entirely constant for quiescent star-forming galaxies.

However, the relationship between $\Sigma_{\textrm{SFR}}$ and $\Sigma_{\textrm{gas}}/\tau_{\textrm{dyn}}$ is at least as tight as the star formation law for our updated sample of spiral and dwarf galaxies, with a correlation coefficient $R=0.80$ close to $R=0.78$ for the Schmidt law (Figure~\ref{fig:KSlawtotal}).
The intrinsic dispersion in the relationship also indicates a tight relation; the linmix estimate of intrinsic scatter is $0.33^{+0.02}_{-0.02}$~dex, slightly smaller than the dispersion estimate for the Schmidt law (intrinsic scatter $0.37^{+0.02}_{-0.02}$~dex).
Yet unlike the Schmidt law, the Silk-Elmegreen law does not have a clear turnover at low gas densities.
This may imply that gas density (the main driving factor in the star formation law) may not be the most important (or only) driver of star formation across different galaxy types.
However, we note that many of the points at low $\Sigma_{\textrm{gas}}/\tau_{\textrm{dyn}}$ are dwarf galaxies with lower limits in $\Sigma_{\textrm{gas}}$, so these points may move to the right and produce a turnover at low gas densities.

We investigate this further by investigating another alternative star formation scaling law: the ``extended Schmidt law'' proposed by \citet{Dopita85} assumes that star formation is affected not only by gas density, but also by the density of existing stars. 
This relationship was originally formulated as
\begin{equation}
\Sigma_{\textrm{SFR}} = A(\Sigma_{\textrm{gas}})^{n}(\Sigma_{*})^{m},
\label{eq:KSext}
\end{equation}
with stellar mass surface density $\Sigma_{*}$. 
Various studies have since found different values of power law indices $n$ and $m$; \citet{Dopita94} initially suggested that $n = 1/3$ and $m = 5/3$. 
More recently, \citet{Shi11} and \citet{Roychowdhury17} found values near $n = 1$ and $m = 0.5$ and showed that with these indices, Equation~\ref{eq:KSext} described a tighter correlation than the Schmidt law. 
Furthermore, on spatially-resolved scales, this extended Schmidt law did not exhibit a threshold and appeared to hold for low surface brightness regions.

\begin{figure}
	\epsscale{1.2}
	\plotone{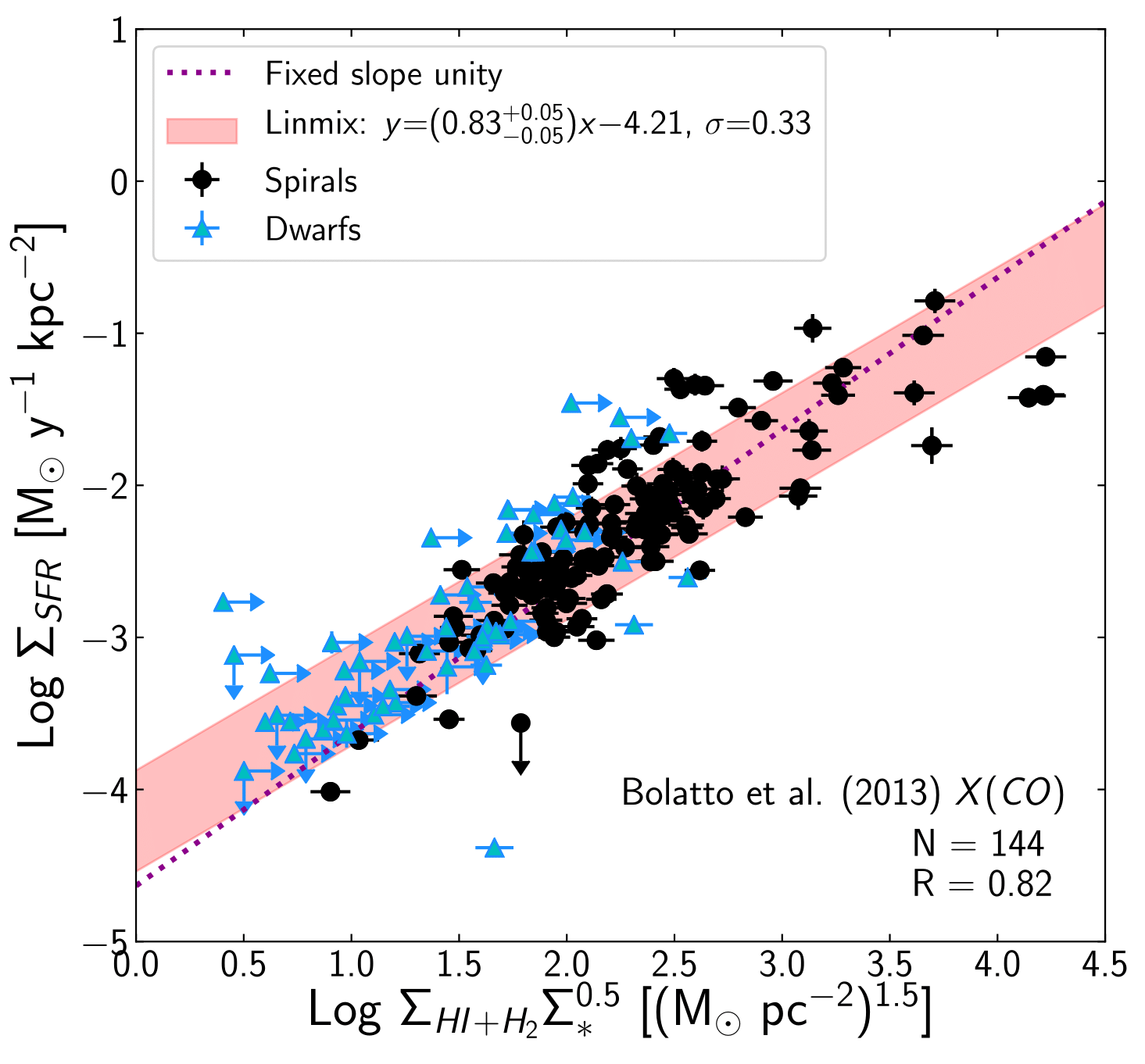}
	\caption{The extended Schmidt relation for spirals (black circles) and dwarfs (cyan triangles), including galaxies with only HI measurements (rightward arrows) and upper limits on SFR measurements (downward arrows). The metallicity-dependent $X(\textrm{CO})$ prescription from \citet{Bolatto13} is used. Red shaded area marks the median linear fit and intrinsic dispersion computed using the linmix MCMC method. The dotted purple line indicates a line of slope unity. \label{fig:KSext}}
\end{figure}

To determine if this result still holds for our sample, we plot the relation between $\Sigma_{\textrm{SFR}}$ and $\Sigma_{\textrm{gas}}\Sigma_{*}^{0.5}$ in Figure~\ref{fig:KSext}.
With a correlation coefficient of $R = 0.82$, this extended Schmidt law is a slightly stronger correlation than the global star formation law ($R=0.80$).
The linmix estimator yields a slightly sublinear slope of $0.83^{+0.05}_{-0.05}$, but the bivariate slope of $1.08\pm0.08$ is consistent with a linear slope as predicted if $n = 1$ and $m = 0.5$. 
However, the extended Schmidt law does not appear to show a turnover at low densities, suggesting that stellar surface densities $\Sigma_{*}$ may be an important parameter in driving star formation in low-$\Sigma_{\textrm{gas}}$ systems.
Again, we note that many of the points at the low-density end are dwarf galaxies with lower limits in $\Sigma_{\textrm{gas}}$, so there may indeed be a low-density threshold.

We return to the importance of $\Sigma_{*}$ later, when we discuss the physical implications of the extended Schmidt law in Section~\ref{sec:interpretations}.

\section{Second-order Correlations} 
\label{sec:secondorder}

The updated SFRs used in this paper are more precise than those used in \citetalias{Kennicutt98}, with measurement uncertainties $\lesssim 0.1$~dex smaller than the uncertainties of $(+0.3,-0.5)$~dex assumed by \citetalias{Kennicutt98}.
However, the scatter in the star formation law (Figure~\ref{fig:KSlawtotal}) is nearly identical ($0.28$~dex in this study, compared to $0.3$~dex).
This suggests that much of the dispersion in the global star formation law is likely intrinsic, an effect further exacerbated when dwarf galaxies are included. 
A correlation between this intrinsic dispersion and a second-order parameter may indicate a more physically ``fundamental'' relationship.
Indeed, Section~\ref{sec:alternativeSFlaws} considers alternative star formation scaling laws that may describe the data better than the canonical star formation law. 
In particular, the ``extended Schmidt law'' suggests that gas density is not the sole driver of star formation, and that stellar density may play a role for dwarf galaxies.
Other secondary parameters could have similarly important implications for the physics of star formation.
We therefore present here a rudimentary test for second-order correlations in the global star formation law.

\begin{figure*}
	\epsscale{1.15}
	\plotone{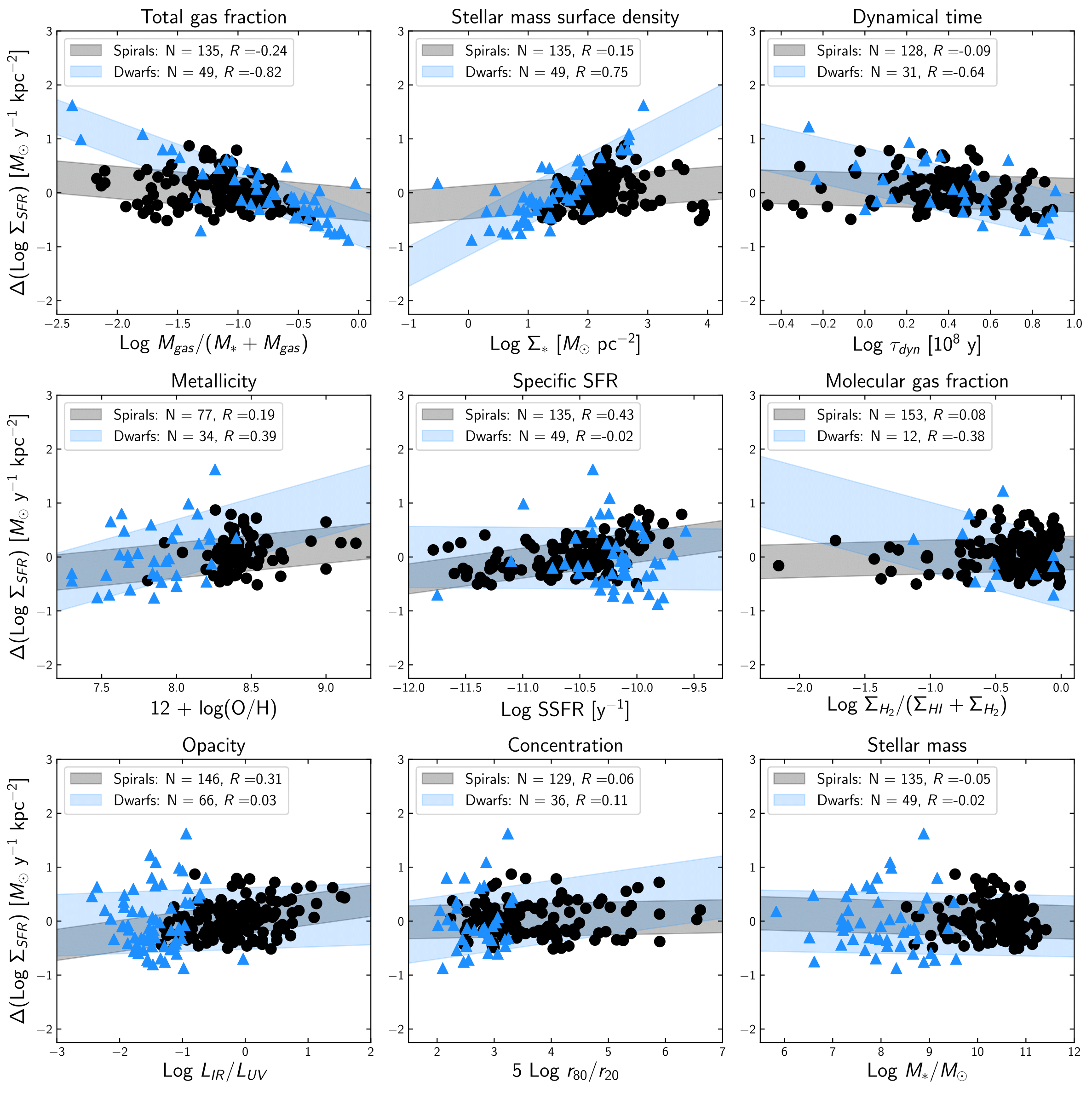}
	\caption{The residuals from the star formation law as a function of various physical parameters. Gray (light blue) shaded regions mark the median linear fits and intrinsic dispersions from the linmix MCMC for spiral (dwarf) galaxies, which are marked as black solid points (cyan triangles). Note that residuals for the dwarf galaxies are computed using just atomic gas, rather than total (atomic + molecular) gas. \label{fig:resids}}
\end{figure*}

We consider the various secondary physical properties described in Section~\ref{sec:otherprops}.
Plotting the scatter in the star formation law as a function of these physical parameters is a simple way to examine which properties show second-order correlations with the star formation law.
We define the scatter in the star formation law as the residuals in the law for the composite dwarf and spiral sample (Eq.~\ref{eq:KSlawfinal}):
\begin{align}
\begin{split}
\Delta\log \Sigma_{\textrm{SFR}} & = \log \Sigma_{\textrm{SFR}_{\textrm{observed}}} - \log \Sigma_{\textrm{SFR}_{\textrm{predicted}}} \\
& = \log \Sigma_{\textrm{SFR}} - (1.26\log \Sigma_{(\textrm{HI}+\textrm{H}_{2})} - 3.78)
\end{split}
\label{eq:resids}
\end{align}

\begin{deluxetable*}{lcccccc}
\tablecolumns{7} 
\tablecaption{ Linear fits to residual plots. \label{tab:resids}} 
\tablehead{ 
\colhead{} & \multicolumn{3}{c}{Spiral galaxies} & \multicolumn{3}{c}{Dwarf galaxies} \\
\colhead{Parameters} & \colhead{$m$} & \colhead{$b$} & \colhead{$\sigma$$^{\textrm{a}}$} & \colhead{$m$} & \colhead{$b$} & \colhead{$\sigma$$^{\textrm{a}}$}}
\startdata
Total gas fraction & $-0.20^{+0.07}_{-0.07}$ & $-0.22^{+0.09}_{-0.08}$ & $0.30^{+0.02}_{-0.02}$ & $-0.82^{+0.09}_{-0.08}$ & $-0.66^{+0.08}_{-0.08}$ & $0.32^{+0.04}_{-0.03}$ \\
Stellar mass surface density & $0.08^{+0.05}_{-0.05}$ & $-0.18^{+0.11}_{-0.11}$ & $0.31^{+0.02}_{-0.02}$ & $0.57^{+0.07}_{-0.07}$ & $-0.80^{+0.11}_{-0.11}$ & $0.37^{+0.04}_{-0.04}$ \\
Dynamical time & $-0.098^{+0.099}_{-0.100}$ & $0.053^{+0.047}_{-0.045}$ & $0.31^{+0.02}_{-0.02}$ & $-0.9^{+0.2}_{-0.2}$ & $0.41^{+0.12}_{-0.12}$ & $0.43^{+0.06}_{-0.05}$ \\
Metallicity & $0.27^{+0.17}_{-0.17}$ & $-2.3^{+1.4}_{-1.5}$ & $0.33^{+0.03}_{-0.03}$ & $0.8^{+0.3}_{-0.3}$ & $-6^{+3}_{-3}$ & $0.54^{+0.08}_{-0.07}$ \\ 
Specific SFR & $0.29^{+0.05}_{-0.05}$ & $3.1^{+0.6}_{-0.6}$ & $0.28^{+0.02}_{-0.02}$ & $-0.02^{+0.18}_{-0.17}$ & $-0.3^{+1.8}_{-1.8}$ & $0.57^{+0.06}_{-0.06}$ \\
Molecular gas fraction & $0.065^{+0.072}_{-0.073}$ & $0.06^{+0.04}_{-0.04}$ & $0.31^{+0.02}_{-0.02}$ & $-0.7^{+0.6}_{-0.6}$ & $-0.30^{+0.34}_{-0.35}$ & $0.7^{+0.2}_{-0.1}$ \\
$L_{\textrm{IR}}/L_{\textrm{UV}}$ & $0.16^{+0.04}_{-0.04}$ & $0.04^{+0.02}_{-0.02}$ & $0.29^{+0.02}_{-0.02}$ & $0.04^{+0.16}_{-0.16}$ & $0.04^{+0.24}_{-0.24}$ & $0.57^{+0.06}_{-0.05}$ \\
Concentration & $0.02^{+0.03}_{-0.03}$ & $-0.06^{+0.11}_{-0.11}$ & $0.31^{+0.02}_{-0.02}$ & $0.16^{+0.25}_{-0.25}$ & $-0.5^{+0.7}_{-0.7}$ & $0.58^{+0.08}_{-0.06}$ \\
Stellar mass & $-0.03^{+0.04}_{-0.05}$ & $0.3^{+0.5}_{-0.5}$ & $0.31^{+0.02}_{-0.02}$ & $-0.013^{+0.096}_{-0.100}$ & $0.08^{+0.80}_{-0.78}$ & $0.57^{+0.07}_{-0.05}$
\enddata
\tablenotetext{a}{As noted in the text, $\sigma$ is the intrinsic dispersion in the $y$-direction estimated by the linmix method (Section~\ref{sec:spirals}).}
\end{deluxetable*}

Figure~\ref{fig:resids} shows plots of these residuals as functions of physical parameters, as well as linmix linear fits to the separate spiral and dwarf samples.
Table~\ref{tab:resids} lists the parameters of these linear fits. 
Since there are so few ($N=12$) dwarf galaxies with both HI and H$_{2}$ measurements, we note that we compute residuals for the dwarf galaxies using only atomic gas (i.e., using $\Sigma_{\textrm{HI}}$ rather than $\Sigma_{(\textrm{HI}+\textrm{H}_{2})}$ in Equation~\ref{eq:resids}).
This may be an appropriate treatment for dwarf galaxies, which are largely dominated by HI gas.

Both Figure~\ref{fig:resids} and Table~\ref{tab:resids} illustrate that for spiral galaxies (black points), the second-order correlations in the star formation law are weak at best.
For example, spiral galaxies show essentially no correlation with molecular gas fraction, with a linear slope consistent with zero slope.
The second-order correlations for the spiral galaxies are particularly weak when compared to the second-order correlations for dwarf galaxies.
Specific SFR and $L_{\textrm{IR}}/L_{\textrm{UV}}$ are the only parameters for which the second-order correlations for spirals are stronger than the correlations for dwarfs.
Spiral galaxies with higher SSFR tend to lie above the fiducial star formation law; this is qualitatively consistent with results from the COLD GASS survey \citep{Saintonge11b}, which found that the molecular gas depletion time for typical star-forming galaxies increases with SSFR.
Similarly, galaxies with higher $L_{\textrm{IR}}/L_{\textrm{UV}}$ tend to lie above the star formation law.
However, dwarf galaxies show essentially no second-order correlation with SSFR or $L_{\textrm{IR}}/L_{\textrm{UV}}$.

For most of the parameters, the dwarf galaxies tend to display stronger second-order correlations than spirals.
Total gas fraction, stellar mass surface density, and dynamical time are perhaps the strongest examples of this phenomenon.
For these parameters, spiral galaxies show only weak correlations between the second-order parameters and residuals ($R=-0.24$ for $f_{\textrm{gas}}$, $R=0.15$ for $\Sigma_{*}$, and $R=-0.09$ for $\tau_{\textrm{dyn}}$).
Dwarf galaxies, however, show much stronger correlations ($R=0.82$, $R=0.75$, and $R=0.64$, respectively).
Furthermore, the linear fits to the dwarf galaxies have steeper slopes than the fits to the spiral galaxies.
For example, when plotting the residuals in the star formation law as a function of total gas fraction, the linmix regression method yields a slope of $-0.82\pm 0.11$ for dwarf galaxies and a smaller slope of $-0.20\pm 0.07$ for spiral galaxies.

Other parameters, including metallicity and molecular gas fraction, show a similar phenomenon to a lesser extent.
While the correlations between these parameters and the star formation law residuals are still weak for spiral galaxies, the correlations are only moderate for dwarf galaxies ($R=0.39$ for metallicity and $R=0.38$ for $f_{\textrm{mol}}$).
For molecular gas fraction, the small sample ($N=12$) of dwarf galaxies with observed H$_{2}$ makes the correlation spurious.
Finally, some parameters---concentration and stellar mass---are weakly correlated with the residuals for both spiral and dwarf galaxies, suggesting that these parameters do not drive any of the scatter in the star formation law.

Which (if any) of these second-order correlations is most physically important? 
That is, can any of the second-order parameters can explain the most scatter in the star formation law?
To more quantitatively assess this, we compare the root-mean-square error (RMSE) of each second-order correlation with the RMSE of the fiducial star formation law (Equation~\ref{eq:KSlawfinal}) for spirals and dwarfs separately.
The RMSE of the fiducial star formation law is $0.31$~dex for spiral galaxies and $0.55$~dex for dwarf galaxies.
We compare these to the RMSE of the second-order correlations, which are estimated by convolving the fiducial star formation law with the linear fits to the residual plots listed in Table~\ref{tab:resids}.

For example, the linear fit to the relation between metallicity and the residuals is $y=0.27x-2.3$ for spiral galaxies, where $x$ is the metallicity $12 + \log[\textrm{O/H}]$ and $y = \Delta\log\Sigma_{\textrm{SFR}}$, defined in Equation~\ref{eq:resids}.
Combining these two equations yields the equation of a plane: 
\begin{equation*}
\log \Sigma_{\textrm{SFR}} = 1.26\log \Sigma_{(\textrm{HI}+\textrm{H}_{2})} + 0.27(12 + \log[\textrm{O/H}]) - 6.08,
\end{equation*}
for which we can then compute the RMSE in $\log \Sigma_{\textrm{SFR}}$.
Repeating this procedure for spiral and dwarf galaxies for each of the parameters, we obtain the RMSEs of the second-order correlations, which are listed in Table~\ref{tab:rmse}.

Table~\ref{tab:rmse} shows that none of the second-order parameters significantly decrease the RMSE for spiral galaxies, implying that the Schmidt law is indeed the most physically relevant star formation scaling law for spiral galaxies.
However, nearly all of the second-order correlations decrease the RMSE for dwarf galaxies.
In particular, total gas fraction, stellar mass surface density, and dynamical time decrease the RMSE for dwarf galaxies by ${\sim}0.1-0.2$~dex, suggesting that these parameters may be important in driving star formation in dwarf galaxies.
We further consider potential physical interpretations in Section~\ref{sec:interpretations}.

We note that we also tried combining the spiral and dwarf galaxy samples, but the combined sample shows only weak second-order correlations with any physical parameters.
None of these correlations significantly decreases the fiducial RMSE.

\begin{deluxetable}{lcc}
\tablecolumns{3} 
\tablecaption{ RMS errors. \label{tab:rmse}} 
\tablehead{ 
\colhead{Parameter} & \colhead{Spirals RMSE} & \colhead{Dwarfs RMSE}\\
\colhead{} & \colhead{(dex)} & \colhead{(dex)}
}
\startdata
Total gas fraction & 0.30 & 0.31 \\
Stellar mass surface density & 0.30 & 0.35 \\
Dynamical time & 0.30 & 0.39 \\
Metallicity & 0.32 & 0.50 \\
Specific SFR & 0.27 & 0.54 \\
Molecular gas fraction & 0.31 & 0.51 \\
$L_{\textrm{IR}}/L_{\textrm{UV}}$ & 0.29 & 0.55 \\
Concentration & 0.31 & 0.54 \\
Stellar mass & 0.31 & 0.54 \\
\textbf{Fiducial SF law} & \textbf{0.31} & \textbf{0.55}
\enddata
\end{deluxetable}

\section{Systematic uncertainties}
\label{sec:uncertainties}
Before considering the implications of our results, we discuss the limitations of our dataset.
Although our dataset represents an improvement to the sample size and measurement uncertainties in the \citetalias{Kennicutt98} dataset, it is still subject to several systematic uncertainties.
We consider some of these here.

\subsection{SFR calibrations}
Estimating star formation rates from multiwavelength tracers is subject to several systematic uncertainties.
We first consider systematics which could potentially influence the overall form and slope of the star formation law, and then mention the main outstanding uncertainties in the zeropoints of the SFR calibrations and the Schmidt law itself.

As discussed in Section~\ref{sec:SFRdata}, in this work we use UV-based SFRs and correct the UV fluxes for dust using the $\sim24\micron{}$ infrared flux. 
The resulting SFR estimates are vastly superior to those presented in \citetalias{Kennicutt98}. 
The UV fluxes average recent star formation over a much longer period than the H$\alpha$ fluxes used in \citetalias{Kennicutt98}; as such, they are much less sensitive to IMF variations and are more reliable in systems with very low SFRs such as quiescent dwarf galaxies, where the IMF is not fully populated in O-type stars.  
In addition, the use of 24\micron{} fluxes to correct each individual UV measurement for dust attenuation is a major improvement over the statistical estimates applied previously.  
The reduction in observational uncertainties in SFRs from of order $\pm0.3$~dex to $\pm0.10$~dex is the main reason for the emergence of a well-defined Schmidt power law in these samples, in comparison to the scatter-dominated correlations for normal galaxies seen in \citetalias{Kennicutt98}.

Nevertheless, some uncertainties remain in these UV$+$IR based SFR measurements.  
As discussed in \citet{Hao11} and \citet{Boquien16}, the relative weighting of UV and IR fluxes in the dust attenuation correction is dependent on the age mix in the dust-heating stellar populations, with the weighting coefficient varying by nearly a factor of two between quiescent populations dominated by older stars and the most actively forming starburst galaxies.  
As discussed in more detail in Paper II, however, the use of the 24\micron{} flux in lieu of the total infrared (TIR) flux mitigates much of this variation, because more active star-forming galaxies have higher 24\micron{}/TIR ratios, largely canceling the deficit in total infrared emission from more evolved stars.  
This systematic effect is probably comparable in magnitude (up to $\pm$0.1 dex) to random errors due to differences in dust geometry, etc.  

Heating of dust by non star-forming stellar populations is another systematic effect that is especially serious in massive, early-type galaxies with very low SFRs.  
Comparisons of SFRs derived using our simple UV$+$IR prescriptions with more detailed SFRs estimated from full SED population synthesis fitting show that in such systems, the empirical recipe can overestimate the UV dust attenuation correction and thus the SFRs by factors of two or more \citep[e.g.,][]{Boquien16, Hunt18}. 
This error has the potential to bias the slope of the derived star formation law, because such galaxies tend to have the lowest surface densities of gas and SFRs in the sample (lower left quadrants of Figures~\ref{fig:KSlawspirals} and \ref{fig:KSatomicmol}). 
In order to constrain the magnitude of any such bias for our sample, we replotted the SFR surface densities for the five spiral galaxies in Figure~\ref{fig:KSlawspirals} with the lowest SFR densities and reddest stellar populations\footnote{These galaxies are NGC 4698, NGC 4216, NGC 0404, NGC 1291, and NGC 5101.}, assuming the extreme case of no dust attenuation.  
Refitting the resulting star formation law steepened the resulting power law slope by $0.04$~dex (from $n=1.41$ to $1.45$).
Note that this is a worst-case example, but it illustrates the need to carefully consider the dust-corrected SFRs when expanding the dynamic range of the sample to those with the lowest gas densities and SFRs.

It is also worth noting that all SFR calibrations are dependent on metallicity.
\citet{Kennicutt12rev} review the theoretical literature and find that the metallicity dependences are relatively modest for UV and IR fluxes; a factor of ten decrease in metal abundance causes roughly a $\sim 0.07\pm 0.03$~dex increase in the FUV and IR luminosities for a fixed SFR.  
The effect is much larger for H$\alpha$ (i.e., ionizing) fluxes, where the increase can be as much as $0.4$~dex over the same range in metallicity.  
Most of the spiral galaxies in our sample span a relatively small range in abundance (a factor of $3-5$ at the extremes), and given the limited information available on abundances for many of the galaxies we have not applied any corrections. 
The dwarf galaxies are of course more metal-poor on average (typically factors of $3-10$, with a few more metal-poor objects).
If we conservatively assume that all dwarf galaxies in our sample are indeed a factor of $10$ more metal-poor than the spirals, we estimate that their SFRs should increase by $\sim 0.07$~dex relative to the spirals.
This metallicity effect does not decrease the slope of the star formation law for the combined sample of dwarf and spiral galaxies.

In summary, although the new SFRs are not entirely free of systematic uncertainties, none that we are aware of are likely to bias the observed slopes of the Schmidt laws by more than 0.1\,dex, comparable to the uncertainties in the fitted slopes themselves.  
As discussed below, we suspect that biases from uncertainties in the CO-derived molecular gas surface densities are far more important.  

We conclude by mentioning sources of systematic error which probably affect all of the SFRs, and thus could bias the zeropoints of the relations. 
The greatest source of uncertainty by far is the form of the stellar IMF, its slope, upper stellar mass limit, and possible systematic variation with metallicity, cluster mass, or star-formation environment. 
Discussion of the IMF falls well beyond the scope of this study but always needs to be borne in mind.  
Likewise, our incomplete understanding of the role of binary stellar evolution and stellar rotation on the luminosities and lifetimes of (especially massive) stars introduce similar global uncertainties into the SFR scales. 
The general consistency of observed galaxy SEDs with models assuming a \citet{Kroupa01} or \citet{Chabrier03} IMF offers some reassurances, but outstanding discrepancies such as the disagreement of SFRs derived from FUV luminosities of galaxies and their resolved color-magnitude diagrams \citep[e.g.,][]{McQuinn15} serves as a reminder that some systematics may not yet be accounted for in existing SFR calibrations.

\subsection{Molecular gas densities}
We have already discussed the potential effects of applying a metallicity-dependent $X(\textrm{CO})$ in some detail in Section~\ref{sec:xco}, particularly focusing on how much $\Sigma_{\textrm{gas}}$ may change for the separate spiral and dwarf samples. 
We found that even for spiral galaxies alone, changing from a constant Milky Way value of $X(\textrm{CO})$ (Equation~\ref{eq:XCO_MW}) to the metallicity-dependent \citet{Bolatto13} $X(\textrm{CO})$ (Equation~\ref{eq:XCO_Bolatto}) decreases the slope of the total gas star formation law from $1.41$ to $1.29$. 
We now consider how much these effects may change the total gas star formation law for the combined dwarf and spiral sample. 

We find that changing $X(\textrm{CO})$ from a constant Milky Way value to the metallicity-dependent \citet{Bolatto13} prescription again decreases the overall slope of the total gas star formation law from $1.47\pm0.08$ to $1.27\pm0.08$. 
Note that this $0.2$ difference in slopes is a conservative estimate of the effect of a metallicity-dependent $X(\textrm{CO})$.
We assume that dwarf galaxies with observed CO are representative of all dwarf galaxies, when these CO-bright dwarfs are likely to be the most strongly affected by changes in $X(\textrm{CO})$.
The effect of the \citet{Bolatto13} $X(\textrm{CO})$ prescription is also less extreme than other prescriptions considered in this paper (see Section~\ref{sec:xco}).
Even this conservative estimate implies that $X(\textrm{CO})$ produces a systematic effect much larger than those arising from the SFR calibration.

\subsection{Choice of diameters}
The surface densities in the star formation law must be scaled by some area computed from (inclination-corrected) diameters. 
Nominally, these diameters represent the star-forming regions in galaxy disks, but obtaining truly representative diameters is often difficult.

\citetalias{Kennicutt98} originally used the optical diameter, defined as the major axis of the $B$-band 25~mag~arcsec$^{-2}$ isophote from the second reference catalog \citep{RC2}.
Other studies have used different scaling diameters, including: the diameters of UV-defined apertures \citep{Wyder09}; the \citet{Holmberg58} size, defined as the $B$-band 26.5~mag~arcsec$^{-2}$ isophote \citep{Roychowdhury14}; diameters computed from resolved $1.4$~GHz radio continuum maps \citep{Liu15}; and the ``near-IR equivalent Holmberg'' size, computed using $(B-H)$ and $(B-K)$ colors \citep{Lopez-Sanchez2018}.

In this work, for reasons discussed in Section~\ref{sec:diams}, we defined the star-forming disk as the region containing ${\sim}95\%$ of the H$\alpha$ flux.
Here, we examine the possibility that this choice of diameters may systematically bias the star formation law.
This effect occurs because changing the diameter of an aperture changes the surface density $\Sigma$ according to the change in flux normalized by the change in area.
However, the radial surface brightness profiles of HI are generally flatter and more extended than the radial profiles of UV or CO fluxes \citep{101}.
As a result, changing the diameter can produce systematically larger changes in $\Sigma_{\textrm{SFR}}$ and $\Sigma_{\textrm{H}_{2}}$ compared to $\Sigma_{\textrm{HI}}$.
Since the fraction of atomic gas is a function of $\Sigma_{\textrm{gas}}$, this can affect the slope of the star formation law.

Here, we demonstrate this effect by considering an extreme case: doubling the diameters for a subset of our galaxies increases the slope of the star formation law by ${\sim}0.2$~dex.
This is a particularly extreme case, especially since our sample is comprised of non-starbursting galaxies, where the star-forming regions are fairly well-defined by our chosen H$\alpha$ diameters.
Nonetheless this example does underscore the importance of choosing appropriate diameters when studying the star formation law.
We discuss this effect in further detail in Paper II, where we consider starburst galaxies in which star formation is confined to a region much smaller than the optical disk.

\section{Comparison with literature}
\label{sec:litcompare}

The primary goal of this investigation has been to update and test the main results from the \citetalias{Kennicutt98} study of the global star formation law, in particular for normal spiral and irregular galaxies.  
This is not the first reassessment of the star formation law since 1998, and although we cite previous works throughout the previous section, it is appropriate to acknowledge the key results in one place and to emphasize that many of the results presented here mainly confirm previous results by other authors.  

The most comprehensive previous study of the integrated star formation law is by \citet{Liu15}.  
They compiled data on gas and SFR surface densities for 115 normal galaxies and 66 luminous and ultraluminous starburst galaxies.  
Since the main focus of that paper is on the combined star formation law for normal and starburst galaxies, we defer most of our discussion of this important study to Paper II.  
However some of their results on normal galaxies can be compared to ours. 
In particular, \citet{Liu15} demonstrate the lack of correlation between disk-averaged SFR surface densities and HI surface densities, and a nearly linear slope of the correlation with H$_{2}$ surface densities, as presented here in Figure~\ref{fig:KSatomicmol}.  
They do not fit the slope of the total gas density relation separately for normal galaxies, but their derived slope $n\sim1.2$ for the combined sample of normal and starburst galaxies appears to be similar to our value of $n=1.27\pm 0.08$ for our combined sample of dwarf and spiral galaxies.
An important difference in their study is that they use 1.4\,GHz radio continuum maps to estimate the SFRs and (along with 24\micron{} fluxes) to measure the SFRs and disk sizes for their sample.
As discussed further in Paper II, the differences in methodology may be able to account for most of the (minor) differences between the results found in that paper and here.

Most papers on the star formation law over the past decade have focused on spatially-resolved measurements of disks, either averaged azimuthally or measured on a point-by-point basis.  
Although there are significant differences in detail between the results of different studies, depending on the methods used (see \citet{Kennicutt12rev} for a review), most show the same decoupling of the SFR surface density with HI and tight, roughly linear dependence on molecular gas surface density \citep[e.g.,][]{Bigiel08,101,225}.
Our new results are in excellent agreement with those results, in marked contrast to \citetalias{Kennicutt98} which found a stronger correlation in disk-averaged surface densities with HI.  
It appears that the differences in conclusions were mainly driven by the small sample sizes in \citetalias{Kennicutt98}, a much less diverse sample of galaxies, and a large variation in $X(\textrm{CO})$ conversion factors in the \citetalias{Kennicutt98} CO sample.  
In any case, this apparent discrepancy between disk-averaged and local Schmidt laws seems to have largely disappeared.

Finally we point out that until recently, a turnover or threshold in the star formation law was only seen in spatially-resolved data \citep[e.g.,][]{Skillman87,Kennicutt89,Martin01,Bigiel08}.
In particular, \citet{101} performed a thorough analysis of this star formation threshold on local scales by examining radial profiles of the star formation efficiency $\Sigma_{\textrm{SFR}}/\Sigma_{\textrm{gas}}$ as a function of various parameters.
On the disk-averaged scale, such a threshold was later observed by \citet{Wyder09} for the disks of low-surface brightness spiral galaxies. 
More recently, \citet{Filho16} and \citet{Roychowdhury17} have observed a similar threshold for gas-rich dwarf galaxies.
\citet{Shi14} also observed a similar feature for a small set of regions in two metal-poor galaxies.  
Our new results significantly strengthen the robustness of these results.

\section{Physical interpretations}
\label{sec:interpretations}

Before discussing the potential implications of our results, we emphasize that star formation is an inherently local process. 
The integrated star formation law averages over huge local variations, making it much more difficult to relate to physical processes.
However, given the stochastic nature of star formation on local scales, we are better equipped to observe the star formation law in a physically meaningful way on disk-averaged scales.
It is therefore worth reviewing physical interpretations that may explain our results.

\subsection{The star formation law for spiral galaxies}

We first consider the shape of the star formation law for spiral galaxies.
With our updated measurements, we found that spiral galaxies alone obey a tight power law with a slope of $n=1.41\pm 0.07$ (assuming a constant value of $X(\textrm{CO})$), consistent with the oft-cited \citetalias{Kennicutt98} value of $n=1.4\pm 0.15$.
Both values of $n$ are conveniently close to $n=1.5$.
This has often been explained by a simplistic argument in which the SFR density scales by the gas density divided by the gravitational free-fall timescale $\tau_{\textrm{ff}}\propto(G\rho_{\textrm{gas}})^{-0.5}$:
\begin{equation} 
\rho_\textrm{SFR}\propto\frac{\rho_{\textrm{gas}}}{\tau_{\textrm{ff}}}\propto\frac{\rho_{\textrm{gas}}}{(G\rho_{\textrm{gas}})^{-0.5}}\propto\rho_{\textrm{gas}}^{1.5}.
\label{eq:KSargument}
\end{equation}
Assuming a constant scale height then yields a Schmidt star formation law with a power law index of $n=1.5$.

On the other hand, the shape of the star formation law has often been explained by spatially-resolved studies as a result of the multi-phase ISM.
The combination of a nearly-linear molecular gas relation and a nearly-vertical atomic gas relation \citep[e.g.,][]{Bigiel08,Leroy13} yields a superlinear slope of $n=1-2$ for the \emph{total} gas star formation law.
To the extent that our results agree with these studies, this could also be a viable explanation for our observed star formation law.

\subsection{Low-density threshold}

Other results from our updated investigation of the global star formation law suggest a more complex picture.
In particular, a low-$\Sigma_{\textrm{gas}}$ threshold in the star formation law appears to separate spiral galaxies from dwarf galaxies and low-surface brightness galaxies.
This is expected in many theoretical frameworks, which predict different star formation regimes for these galaxies. 

For example, this turnover has been observed in the star formation law within radial profiles of disk galaxies \citep[e.g.,][]{Kennicutt89}.
Early explanations for this threshold invoked the Toomre-Q criterion for gas stability, which defines some critical density below which gas is stable against collapse, suppressing star formation.
The same self-gravitational framework may also explain the threshold using differences in scale height.
In spiral galaxies, the scale height is roughly constant, so that the volumetric densities $\rho$ in Equation~\ref{eq:KSargument} can be converted to surface densities $\Sigma$, leading to a Schmidt power law with index $n\sim1.5$.
However, in dwarf galaxies or flaring disks, the gas scale height might be inversely proportional to $\Sigma_{\textrm{gas}}$.
This leads to a Schmidt power law with $n\sim2$, which is seen in multiple spatially-resolved studies of dwarf irregulars \citep[e.g.,][]{Ferguson98,Elmegreen15}.
We are unable to check these predictions, since we cannot fit a slope in the low-$\Sigma_{\textrm{gas}}$ regime with any certainty given the small sample size and large systematic uncertainties in this regime (Section~\ref{sec:uncertainties}).
Observational work to account for variations in scale height by studying the volumetric star formation law (i.e., $\rho_{\textrm{SFR}}$ vs. $\rho_{\textrm{gas}}$) is now ongoing \citep[][K. Yim, priv. comm.]{Bacchini18}.

More recent studies on local scales have considered another explanation, again suggesting that the low-density threshold may simply result from the multi-phase ISM.
In this picture, the low-density turnover is a phase transition below which the ISM becomes predominantly atomic gas, producing a steep, near-vertical star formation law \citep[e.g.,][]{101, Leroy13, Bigiel14}.
Both the gravitational stability and phase transition models may explain our observed low-$\Sigma_{\textrm{gas}}$ threshold.

Various authors have also attempted to find alternative scaling laws that remove the threshold entirely.
We find that one of these, the ``extended Schmidt law'' \citep{Dopita85,Shi11,Roychowdhury17}, is indeed a marginally better fit than the conventional Schmidt law for the combined sample of spiral and dwarf galaxies (Section~\ref{sec:alternativeSFlaws}).
This law, which sets $\Sigma_{\textrm{SFR}}\propto\Sigma_{\textrm{gas}}\Sigma_{*}^{0.5}$, may physically arise from stellar feedback regulating star formation \citep[e.g.,][]{Orr18}. 
In this picture, young massive stars inject pressure into the interstellar medium, so that this feedback pressure is roughly proportional to SFR density. 
For a system in equilibrium, this pressure must be balanced by hydrostatic pressure. 
The midplane hydrostatic pressure can be written as a combination of the pressure of gas in the stellar potential (proportional to $\Sigma_{\textrm{gas}}\Sigma_{*}^{0.5}$) and gas self-gravity (proportional to $\Sigma_{\textrm{gas}}^{1.5}$) \citep[e.g.,][]{Blitz04, Kim11, Kim15}.
The stellar potential term $\Sigma_{\textrm{gas}}\Sigma_{*}^{0.5}$ becomes significant in low-$\Sigma_{\textrm{gas}}$ systems like dwarf galaxies.
The extended Schmidt law may therefore better describe low-$\Sigma_{\textrm{gas}}$ galaxies, eliminating the low-density turnover observed in the star formation law.

\subsection{Second-order correlations}

In Section~\ref{sec:secondorder} we investigated potential second-order correlations in the star formation law. 
We found that spiral galaxies tended to display weak second-order correlations with other galactic properties.
This is somewhat surprising, given that previous works have suggested that parameters like molecular gas fraction and stellar mass surface density should explain many of the features in the star formation law.

On the other hand, the star formation law for dwarf galaxies tended to exhibit stronger second-order correlations with other galactic parameters.
To determine which of these correlations could explain the scatter in the star formation law, for each parameter $X$ we compared the RMS error in the $\Sigma_{\textrm{SFR}}$-$\Sigma_{\textrm{gas}}$-$X$ plane to the RMS error in the canonical star formation law.
We note that this is not a full multivariate analysis, but it should be sufficient to discover any obvious trends.
We found that for dwarf galaxies, the correlations with total gas fraction, stellar mass surface density, and dynamical time most strongly decreased the RMSE in the star formation law (Table~\ref{tab:rmse}).

The parameter that most significantly decreases the scatter in the star formation law for dwarf galaxies is total gas fraction $f_{\textrm{gas}} = M_{\textrm{gas}}/(M_{\textrm{gas}}+M_{*})$.
The second-order correlation with $f_{\textrm{gas}}$ suggests that at a given $\Sigma_{\textrm{gas}}$, total gas fraction $f_{\textrm{gas}}$ increases.
Both gas mass and stellar mass are measured roughly within the same star forming region, so $f_{\textrm{gas}} \approx \Sigma_{\textrm{gas}}/(\Sigma_{\textrm{gas}}+\Sigma_{*})$.
This means that at a given $\Sigma_{\textrm{gas}}$, the stellar mass surface density $\Sigma_{*}$ increases as $\Sigma_{\textrm{SFR}}$ increases.
Indeed, we do see this second-order correlation with $\Sigma_{*}$. 
Both of these may simply be a restatement of the extended Schmidt law; as discussed in the previous subsection, this may arise from a feedback-regulated model of star formation.

Alternatively, \citet{Shi11} also point out that the extended Schmidt law may arise from the effects of metallicity.
This occurs because $\Sigma_{*}$ traces the total metal enrichment in a galaxy, which is roughly correlated with the gas-phase metallicity.
We do see a second-order correlation between the residuals of the star formation law and metallicity, but metallicity appears to directly explain less of the scatter in the star formation law than other parameters (cf. Table~\ref{tab:rmse}).

The correlation with dynamical time suggests that for dwarf galaxies at a given $\Sigma_{\textrm{gas}}$, dynamical time decreases as $\Sigma_{\textrm{SFR}}$ increases.
This may imply that global processes are particularly important for star formation in dwarf galaxies, since a shorter $\tau_{\textrm{dyn}}$ implies more rapid global dynamical processes that induce faster SFRs. 
This picture of globally-induced star formation could also explain the extended Schmidt law.
Since $\tau_{\textrm{dyn}}$ is defined as the orbital timescale at the radius of the star-forming region, it is inversely proportional to the mass interior to this radius. 
Decreasing the dynamical time therefore increases the mass interior to the star-forming radius; for a given $\Sigma_{\textrm{gas}}$, this increases $\Sigma_{*}$, producing the dependence on $\Sigma_{*}$ seen in the extended Schmidt law.
However, it is unclear why these global dynamical processes might be more important in dwarf galaxies than in spiral galaxies.

All of these second-order correlations are potentially subject to the systematic uncertainties listed in Section~\ref{sec:uncertainties}, as well as uncertainties inherent in our fitting techniques and small sample of dwarf galaxies. 
As a result, the interpretations offered in this section are merely potential interpretations, and we strongly caution against making conclusive quantitative statements about these second-order correlations.

\section{Summary}
\label{sec:summary}
Twenty years after the work of \citet{Kennicutt98}, we have revisited the global star formation law with an improved sample of local star-forming spiral and dwarf galaxies.
In general, we find that the commonly-used $n\sim1.4$ power law from \citetalias{Kennicutt98} is still a reasonable approximation for non-starbursting galaxies.
However, the physics behind the star formation law remain unclear, and we urge the reader to keep in mind that it comes with many caveats.
We now summarize our major results here.

\begin{enumerate}
\item We have confirmed that spiral galaxies alone obey a tight correlation between gas and star formation rate surface densities (Section~\ref{sec:spirals}):
\begin{equation*}
\log\Sigma_{\textrm{SFR}} = 1.41\log\Sigma_{\textrm{gas}} - 3.74,
\end{equation*} 
where $\Sigma_{\textrm{gas}}$ is the sum of both molecular and atomic hydrogen gas surface densities.
We note that starburst galaxies are no longer necessary to define this tight star formation law, as they were in \citetalias{Kennicutt98}. 

\item We found that for spiral galaxies, $\Sigma_{\textrm{SFR}}$ is only weakly dependent on HI gas surface density but scales roughly linearly with H$_{2}$ gas surface density (Section~\ref{sec:molatomSFlaws}).
This is more consistent with what is seen in spatially-resolved studies of the star formation law.

\item We extended the star formation law to include dwarf galaxies using self-consistent measurement techniques. 
Although H$_{2}$ surface densities are heavily dependent on the $X(\textrm{CO})$ conversion factor, we found that dwarf galaxies tend to fall below the star formation law for spirals, producing a turnover in the law at low $\Sigma_{\textrm{gas}}$ (Section~\ref{sec:spiraldwarfs}).

\item We also considered alternative star formation scaling laws (Section~\ref{sec:alternativeSFlaws}), including the relation between $\Sigma_{\textrm{SFR}}$ and $\Sigma_{\textrm{gas}}/\tau_{\textrm{dyn}}$ (the Silk-Elmegreen relation) and the relation between $\Sigma_{\textrm{SFR}}$ and $\Sigma_{\textrm{gas}}\Sigma_{*}^{0.5}$ (the extended Schmidt law).
We found that while both relations were as strongly correlated as the Schmidt star formation law, the extended Schmidt law removed much of the low-$\Sigma_{\textrm{gas}}$ threshold.

\item We found that much of the scatter in the star formation law is intrinsic, motivating a search for second-order correlations in the star formation law (Section~\ref{sec:secondorder}). 
We found that there are no significant second-order correlations for spiral galaxies, but that second-order correlations with total gas fraction ($\frac{M_{\textrm{gas}}}{M_{\textrm{gas}}+M_{*}}$), $\Sigma_{*}$, or $\tau_{\textrm{dyn}}$ may explain much of the scatter in the star formation law for dwarf galaxies.

\end{enumerate}

Again, we note that there are several systematic uncertainties that affect these results, particularly the choice of diameter of the star-forming region and the $X(\textrm{CO})$ factor (Section~\ref{sec:uncertainties}).
These uncertainties should be borne in mind when interpreting our results; we offer some potential physical interpretations nonetheless (Section~\ref{sec:interpretations}).
In the future, we aim to extend this work to revisit local circumnuclear starburst galaxies in Paper II (Kennicutt \& de los Reyes, in prep.). 
Other future work could more carefully consider the systematic effects discussed in this work, as well as revisit the simple statistical analysis used to determine second-order effects.

\acknowledgments
This research has made use of the NASA/IPAC Extragalactic Database (NED), which is operated by the Jet Propulsion Laboratory, California Institute of Technology, under contract with the National Aeronautics and Space Administration.
This research was supported in part by the STFC through a consolidated grant to the Institute of Astronomy, University of Cambridge.
M. A. de los Reyes also acknowledges the financial support of the Winston Churchill Foundation and the NSF Graduate Research Fellowship Program.

The authors would like to thank the anonymous referee for their thoughtful and constructive comments, as well as M. Irwin, A. Saintonge, L. Hunt, and J. Wang for their useful comments and advice.
Finally, we would like to express our deep gratitude to the staff at academic and telescope facilities, particularly those whose communities are excluded from the academic system, but whose labor maintains spaces for scientific inquiry.

\software{
Matplotlib \citep{matplotlib}, 
Linmix \citep{linmix}, 
Astropy \citep{astropy}
}

\bibliographystyle{aasjournal}
\bibliography{Astro-galaxies-KS}

\appendix
\section{Photometry Procedure and Corrections}

\subsection{UV and IR photometry}
\label{appendix:photometry}
This appendix describes the aperture photometry methods we used to obtain UV and IR fluxes for galaxies without available literature data.
The photometry procedures are similar to those described in \citet{lee11} and \citet{dale09} for UV and IR imaging, respectively, of galaxies in the Local Volume Legacy.
As discussed in the text, we measured fluxes within elliptical apertures containing ${\sim}95\%$ of the H$\alpha$ flux.

\textbf{Contaminant removal.}
We first masked potential contaminants such as foreground stars and background galaxies. 
We used the \texttt{irafstarfind}\footnote{\url{http://photutils.readthedocs.io/en/v0.2/api/photutils.irafstarfind.html}} function from the Python package photutils \citep[now an affiliated package of the Astropy library; see][]{astropy} to identify all point sources.
For UV images, any sources with $(m_{\textrm{FUV}} - m_{\textrm{NUV}}) > 1$ were masked as foreground Galactic stars.
These preliminary masks were then inspected by eye and compared against optical images from the Hubble Legacy Archive and the Sloan Digital Sky Survey.
For the Spitzer \mips{} and AllWISE mid-IR images, we masked obvious diffraction spikes and background galaxies identified in the archival optical images.
The masked pixels were linearly interpolated from the surrounding pixels, a negligible correction: the average percent difference between corrected and uncorrected flux was $\sim0.01\%$ for \galex{} UV photometry, $\sim0.11\%$ for mid-IR photometry using AllWISE images, and $\sim0.28\%$ for mid-IR photometry using Spitzer \mips{} images.

\textbf{Background subtraction.}
Constant backgrounds were already subtracted from the mid-IR Spitzer \mips{} images during pre-processing.
For the \galex{} UV images and mid-IR images from ALLWISE, the sky background level was calculated using an image completely masked of all objects in the field, including the target galaxy. 
A circular annulus centered around the location of the target galaxy was divided into $100$ equal-area regions. 
The annulus had a typical inner radius of $2R_{25}$ (except in rare cases where this radius extended outside of the \galex{} field of view; in these cases, the galaxy's emission often appeared sufficiently compact such that a smaller inner radius of $1.25R_{25}$ could be safely used) and an outer radius 75~arcsec beyond the inner radius.
Here $R_{25}$ is the semimajor axis of the $B$-band 25~mag~arcsec$^{-2}$ isophote ($R_{25}=D_{25}/2$).
From the equal-area subregions of the annulus, we computed the average sky background and standard deviation per pixel.
We subtracted this constant sky background from the contaminant-masked image to produce the final image.

\textbf{Aperture photometry.}
We then defined an elliptical aperture with a semimajor axis $a$ given by the diameter of the H$\alpha$-defined star-forming region.
Both the axis ratio $b/a$ and the position angle $\theta$ of the ellipse were determined from \citetalias{RC3}.
Each aperture was inspected by eye; the final semimajor axes and position angle are listed in Table~\ref{tab:gendata}.
Finally, the photutils Python package \texttt{aperture\_photometry} was used to sum the intensity units within the elliptical aperture, resulting in the sum $C_{\textrm{sky-subtracted}}$. 

For \galex{} UV images, the intensity units are given by counts/s. 
Measurement uncertainty (in counts/s) is given by
\begin{equation}
\label{eq:UVuncert}
\sigma = \sqrt{(\sqrt{C_{\textrm{sky-subtracted}}})^{2} + n(\sigma_{sky})^{2}}.
\end{equation}
Here, the first term represents the Poisson counting error, while the second term is the uncertainty in sky background; $n$ is the number of pixels within the aperture and $\sigma_{sky}$ is the measured standard deviation of the sky per pixel.
Finally, both $C_{\textrm{sky-subtracted}}$ and $\sigma$ were then converted to an AB magnitude and error \citep{oke90} according to the \galex{} prescription:
\begin{equation*}
m_{\textrm{FUV}} = -2.5 \times \log_{10}(C_{\textrm{sky-subtracted}}) + 18.82.
\end{equation*}

Following the procedure of \citet{lee11}, this magnitude was then corrected for Galactic extinction using the following formula:
\begin{equation*}
m_{\textrm{FUV, corrected}} = m_{\textrm{FUV, measured}} - A_{\textrm{FUV}}
\end{equation*}
where $A_{\textrm{FUV}} = 7.9E(B-V)$, assuming a total-to-selective extinction ratio of $R_{V}=3.1$ \citep{cardelli89}.
The reddening $E(B-V)$, which is reported in Table~\ref{tab:gendata}, was given by the dust maps of \citet{schlegel98}.
The only exception is IC 0010, which is located close to the Galactic plane where \citet{schlegel98} values become unreliable; for this galaxy, we adopt the independently measured value of $E(B-V) = 0.77\pm 0.07$ from \citet{Richer01}.

The total UV photometric uncertainty is the sum in quadrature of: the measurement uncertainty (Eq.~\ref{eq:UVuncert}), the uncertainty in $E(B-V)$, and the absolute calibration uncertainty in FUV (0.05 mag) or NUV (0.03 mag) for \galex{} \citep{Hao11}.

For Spitzer \mips{} images, which have intensity units of MJy/sr, total photometric uncertainty (in MJy/sr) is given by 
\begin{equation*}
\label{eq:Spitzeruncert}
\sigma = \sqrt{(\sqrt{C_{\textrm{sky-subtracted}}})^{2} + \sigma_{\textrm{calibration}}^{2}}.
\end{equation*}
Again, the first term represents Poisson counting error.
Since the images are already background-subtracted, the second term $\sigma_{\textrm{calibration}}$ is simply the 4\% \mips{} calibration uncertainty at 24~$\mu$m \citep[see, e.g.,][]{engelbracht07}. 
Using the pixel resolution of 24~$\mu$m \mips{} images (1.5''), $C_{\textrm{sky-subtracted}}$ and $\sigma$ were converted from MJy/sr to Jy/pixel, then multiplied by the number of pixels within the aperture $n$ to get total aperture flux and uncertainty in Jy.

For AllWISE images, which have intensity units of digital numbers (DN), the $1\sigma$ uncertainty in DN is given in the AllWISE Explanatory Supplement as:
\begin{equation*}
\sigma = \sqrt{F_\textrm{corr}\left(\sum_{i}^{n}\sigma_{i}^{2} + \frac{n^{2}}{n_{\textrm{annulus}}}\sigma_{sky}^{2}\right)}.
\end{equation*}

\begin{figure}
	\epsscale{0.5}
	\plotone{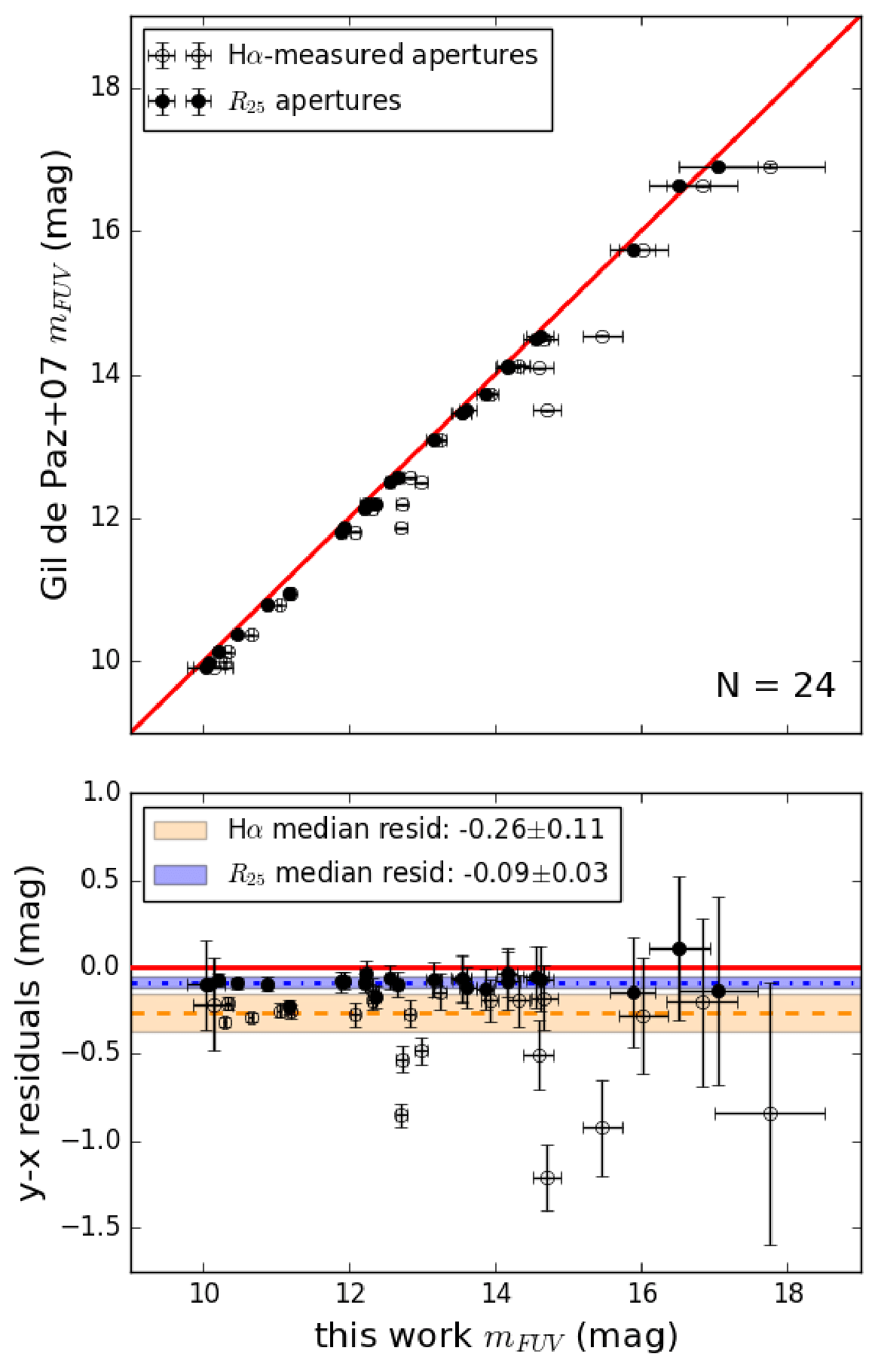}
	\caption{Comparison between our measured FUV fluxes and the FUV fluxes reported in \citet{GildePaz2007}. The empty points denote our measurements using H$\alpha$-based apertures, while the filled points denote our measurements using $R_{25}$-based apertures. (Top) Direct comparison, with the 1-1 line shown in solid red. (Bottom) The $y-x$ residuals, with zero as a solid red line. The median difference between our H$\alpha$-aperture ($R_{25}$-aperture) measurements and the \citet{GildePaz2007} fluxes is denoted by a dot-dashed orange (dashed blue) line; the orange (blue) shaded region marks $\pm 1$ median absolute deviation. \label{fig:UVaperture}}
\end{figure}

Here, $F_\textrm{corr}$ is the correlated noise correction factor for flux variance, given by $F_\textrm{corr} \sim 391.392$ for the 22~$\mu$m W4 band.
The term $\sum_{i}^{n}\sigma_{i}^{2}$ is the sum of flux uncertainties (given by the uncertainty map) for all pixels within the aperture.
The number of pixels within the source aperture and the number of pixels within the background annulus are denoted by $n$ and $n_{\textrm{annulus}}$, respectively.
Since the local galaxies in the sample are largely bright and extended sources, we assumed any confusion noise from faint and unresolved sources is negligible.
Finally, both $C_{\textrm{sky-subtracted}}$ and $\sigma$ could be converted from units of DN to Jy using calibration constants listed in the AllWISE Explanatory Supplement:
\begin{align*}
F~\textrm{(Jy)} & = (5.2269\times 10^{-5})C_{\textrm{sky-subtracted}} \\
\sigma_{F}~\textrm{(Jy)} & = (5.2269\times 10^{-5})\sqrt{(3.4622\times 10^{-4})C_{\textrm{sky-subtracted}}^{2} + \sigma^{2}}.
\end{align*}

\subsection{Photometry systematics and corrections}
\label{appendix:photcorrections}
To ensure that our photometric flux measurements are consistent with existing catalogs, we compared against catalog fluxes.

As Figure~\ref{fig:UVaperture} shows, the FUV fluxes we measured within H$\alpha$-defined star forming regions (empty points) were consistently smaller than the catalog fluxes.
This discrepancy decreases significantly when the photometry is repeated using apertures with larger semi-major axes determined by $R_{25}$ (filled points).
This suggests that the discrepancy is due to systematic differences in aperture size, since most catalogs report estimates of total fluxes rather than aperture fluxes.
Figure~\ref{fig:UVaperture} only shows the comparison with \citet{GildePaz2007} UV fluxes, but a similar effect occurs for the other UV catalogs.

\begin{table}
\caption{Correction factors for UV and IR catalogs.}
\centering
\label{tab:apcorr}
\begin{tabular}{l l}
\hline 
\multicolumn{2}{c}{UV catalogs} \\
{Catalog} & {$f_{corr}$} \\ 
\hline
\citet{GildePaz2007} & $1.27\pm 0.13$ \\
LVL, using IR-matched apertures \citep{lee11} & $1.36\pm 0.15$ \\
SINGS \citep{dale07} & $1.31\pm 0.17$ \\
LVL, using ``outermost elliptical annulus'' apertures \citep{lee11} & $1.49\pm 0.29$ \\
\citet{Bai15} & $1.14\pm 0.12$ \\
Virgo Cluster Survey \citep{Voyer14} & $0.98\pm 0.07$ \\
Herschel Reference Survey \citep{Cortese12} & $0.94\pm 0.10$ \\
\hline
\multicolumn{2}{c}{IR catalogs} \\
{Catalog} & {$f_{corr}$} \\ 
\hline
LVL \citep{dale09} & $1.08\pm 0.07$ \\
SINGS \citep{dale07} & $1.07\pm 0.07$ \\
\citet{GildePaz2007} & $1.10\pm 0.20$ \\
\mips{} LG \citep{Bendo12} & $1.11\pm 0.08$ \\
\iras{} BGS \citep{Sanders03} & $1.19\pm 0.25$ \\
AllWISE measurements & $1.07\pm 0.08$ \\
\hline
\end{tabular}
\end{table}

To account for these systematic effects in the UV, we applied a statistical correction to the catalog FUV fluxes, determined from the median difference between our measured fluxes (using H$\alpha$-based apertures) and the fluxes from each catalog.
For example, the \citet{GildePaz2007} UV fluxes are brighter than our measured fluxes by a median difference of $0.26\pm 0.11$~mag; we therefore divided all \citet{GildePaz2007} UV fluxes by a corresponding factor of $f_{\textrm{corr}} = 1.27\pm 0.13$.
The dispersion $\sigma_{\textrm{corr}} = 0.13$ was added in quadrature with the photometric uncertainty to produce the final flux uncertainties.

A similar aperture effect occurs with the IR photometric fluxes. 
Furthermore, the mid-IR fluxes were not all measured at the same wavelengths: Spitzer \mips{} measures fluxes at 24~$\mu$m, while \iras{} measures fluxes at 25~$\mu$m and \wise{} measures 22~$\mu$m.
We therefore compute similar statistical corrections to calibrate all IR catalogs to agree with the fluxes measured from Spitzer \mips{} images.

Finally, the IR fluxes we measured from AllWISE images were also higher than the fluxes we measured from Spitzer \mips{} images, although the the same apertures were used for both.
This discrepancy may again be due to the effects of different wavelengths (Spitzer \mips{} at 24~$\mu$m, AllWISE at 22~$\mu$m).
It may also result from different background subtraction techniques.
The Spitzer \mips{} images used for photometry were obtained from the SINGS, LVL, and \mips{} LG surveys, which each used some form of polynomial fitting to estimate background\footnote{See SINGS documentation at \url{http://irsa.ipac.caltech.edu/data/SPITZER/SINGS/doc/sings\_fifth\_delivery\_v2.pdf}, LVL documentation at \url{http://irsa.ipac.caltech.edu/data/SPITZER/LVL/LVL\_DR5\_v5.pdf}, and \mips{} LG documentation in \citet{Bendo12}.}.
For the AllWISE images, on the other hand, only a constant average background value was subtracted.
The polynomial fitting used on the \mips{} images likely produced higher interpolated background values at the locations of the target galaxies, producing the discrepancy between \mips{} and AllWISE fluxes.
Using the median difference between the \mips{} and AllWISE fluxes, we applied another statistical correction to the AllWISE fluxes to account for this discrepancy.

The final correction factors for all UV and IR catalogs are listed in Table~\ref{tab:apcorr}. 
Tables~\ref{tab:UVdata} and \ref{tab:IRdata} list the final aperture-corrected UV and IR fluxes that are used in this work.

\section{References for gas surface densities}
\label{appendix:gasrefs}
We list the references for Table~\ref{tab:photometry} here.
The numbering scheme largely follows a convention: references 1-21 contain both HI and H$_{2}$ data, references 101-157 contain primarily HI data, and references 201-241 contain primarily H$_{2}$ data.

1: \citet{1};
2: \citet{2};
3: \citet{3};
4: \citet{4};
5: \citet{5};
6: \citet{6};
7: \citet{7};
8: \citet{8};
9: \citet{9};
10: \citet{10};
11: \citet{11};
12: \citet{12};
13: \citet{13};
14: \citet{14};
15: \citet{15};
16: \citet{16};
17: \citet{17};
18: \citet{18};
19: \citet{19};
20: \citet{20};
21: \citet{21}

101: \citet{101};
102: \citet{102};
103: \citet{103};
104: \citet{104};
105: \citet{105};
106: \citet{106};
107: \citet{107};
108: \citet{108};
109: \citet{109};
110: \citet{110};
111: \citet{111};
112: \citet{112};
113: \citet{113};
114: \citet{114};
115: \citet{115};
116: \citet{116};
117: \citet{117};
118: \citet{118};
119: \citet{119};
120: \citet{120};
121: \citet{121};
122: \citet{122};
123: \citet{123};
124: \citet{124};
125: \citet{125};
126: \citet{126};
127: \citet{127};
128: \citet{128};
129: \citet{129};
130: \citet{130};
131: \citet{131};
132: \citet{132};
133: \citet{133};
134: \citet{134};
135: \citet{135};
136: \citet{136};
137: \citet{137};
138: \citet{138};
139: \citet{139};
140: \citet{140};
141: \citet{141};
142: \citet{142};
143: \citet{143};
144: \citet{144};
145: \citet{145};
146: \citet{146};
147: \citet{147};
148: \citet{148};
149: \citet{149};
150: \citet{150};
151: \citet{151};
152: \citet{152};
153: \citet{153};
154: \citet{154};
155: \citet{155};
156: \citet{156};
157: \citet{157}

201: \citet{201};
202: \citet{202};
203: \citet{203};
204: \citet{204};
205: \citet{205};
206: \citet{206};
207: \citet{207};
208: \citet{208};
209: \citet{209};
210: \citet{210};
211: \citet{211};
212: \citet{212};
213: \citet{213};
214: \citet{214};
215: \citet{215};
216: \citet{216};
217: \citet{217};
218: \citet{218};
219: \citet{219};
220: \citet{220};
221: \citet{221};
222: \citet{222};
223: \citet{223};
224: \citet{224};
225: \citet{225};
226: \citet{226};
227: \citet{227};
228: \citet{228};
229: \citet{229};
230: \citet{230};
231: \citet{231};
232: \citet{232};
233: \citet{233};
234: \citet{234};
235: \citet{235};
236: \citet{236};
237: \citet{237};
238: \citet{238};
239: \citet{239};
240: \citet{240};
241: \citet{241}

\end{document}